\documentclass[aps,prl,reprint,amsfonts,amsmath,amssymb,longbibliography,showkeys]{revtex4-2}
\usepackage{hyperref}
\usepackage{graphicx}
\usepackage{bm}
\usepackage{newtxtext}
\usepackage[varg]{newtxmath}
\usepackage[normalem]{ulem}
\usepackage{braket}
\hypersetup{colorlinks=true,linkcolor=blue,citecolor=blue,urlcolor=blue}
\usepackage{mathtools}
\usepackage{cancel}
\usepackage{mleftright}
\mleftright

\DeclareMathOperator*{\Tr}{Tr}

\DeclareMathOperator{\re}{Re}

\begin{document}

\title{Extracting Nonlinear Dynamical Response Functions from Time Evolution}
\author{\mbox{Atsushi Ono}}
\affiliation{\mbox{Department of Physics, Tohoku University, Sendai 980-8578, Japan}}
\date{July 9, 2025}

\begin{abstract}
We develop a general framework based on the functional derivative to extract nonlinear dynamical response functions from the temporal evolution of physical quantities, without explicitly computing multipoint correlation functions.
We validate our approach by calculating the second- and third-order optical responses in the Rice--Mele model and further apply it to a many-body interacting system using a tensor network method.
This framework is broadly applicable to any method that can compute real-time dynamics, offering a powerful and versatile tool for investigating nonlinear responses in dynamical systems.
\end{abstract}

\makeatletter
\renewcommand{\@keys@name}{DOI: }
\makeatother
\keywords{\href{https://doi.org/10.1103/qsxr-c2pq}{10.1103/qsxr-c2pq}}

\maketitle

\pdfbookmark[1]{Introduction}{introduction}\textit{Introduction}---%
Nonlinear responses have become a cornerstone of modern physics, uncovering a wealth of intriguing and often unexpected phenomena in a wide variety of dynamical systems \cite{Boyd2020, Shen2002, Liu2022, Chang2014, Dorfman2016, Kauranen2012, Grifoni1998}.
Over the decades, nonlinear phenomena in solids, such as harmonic generation \cite{Ghimire2019, Goulielmakis2022, Park2022} and nonlinear Hall effects \cite{Du2021, Bandyopadhyay2024}, have attracted growing interest---not only for their practical applications in technology but also for their profound connections to the geometrical and topological properties of materials \cite{Orenstein2021, Bao2021, Ma2021, Morimoto2023, Ahn2020, Ahn2022, Bhalla2022, Watanabe2021}.
Central to these investigations are nonlinear response functions, which provide the theoretical foundation for understanding such phenomena.
These functions, represented as multipoint correlation functions, are most often calculated by using perturbative expressions in the frequency domain, leveraging frameworks such as the density-matrix formalism \cite{Sipe1993, Sipe2000, Ventura2017, Passos2018, Watanabe2021} and Green's function or Feynman diagram techniques \cite{Tsuji2015, DeJuan2017, Parker2019a, Holder2020, Joao2020, Tsuji2016, Tsuji2020, Tanabe2021, Oiwa2022, Huang2023, Brenig2024, Michishita2021, Chang2024}.

Studies based on real-time evolution have also been conducted to obtain response functions in the frequency domain.
A key advantage of real-time evolution methods is their ability to incorporate various effects.
For instance, tensor network methods accurately account for many-body correlation effects, the quantum master equation includes dissipation effects, and mean-field dynamics capture certain types of vertex corrections.
To date, linear responses have been widely studied through time evolution simulations \cite{Bertsch2000, White2004, Pereira2009, DiasdaSilva2008, Karrasch2013, Phien2012, VanDamme2021, Lange2018d, Yang2019q, Takayoshi2018, Ohmura2019, Murakami2021, Kanega2021, Udono2022, Gohlke2017, Nasu2019, Misawa2023, Cookmeyer2023, Kato2025, Kokcu2024, AriasEspinoza2024, Murakami2020b, Kaneko2021, Iguchi2024, Hattori2025, Hattori2024, Shao2016}, and this approach has also been extended to nonequilibrium transient responses \cite{Eckstein2008, DeFilippis2012, Lenarcic2014, Shao2016, Shinjo2018, Kumar2019, Eskandari-asl2024, Ejima2022, Udono2023, Tohyama2023, Takubo2024, Shinjo2024, Shinjo2025, Osterkorn2024, Kim2024}.
In contrast, studies on nonlinear responses remain relatively scarce.
The methods employed in these studies can be broadly categorized into two approaches: (i) single-pulse excitation and (ii) multiple-pulse excitation.

In the (i) single-pulse excitation method, a monochromatic pulse or continuous wave is applied, and nonlinear responses are extracted by analyzing their amplitude and frequency dependence \cite{Attaccalite2013, Uemoto2019, Kaneko2021, Skachkov2024, Iguchi2024, Hattori2025, Kofuji2024}.
This approach is well suited for addressing specific nonlinear phenomena, such as the second and third harmonic generation, and the optical rectification (including photovoltaic effects).
However, this method provides only \textit{partial} nonlinear response functions as they are limited to specific points in frequency space.

A representative example of the (ii) multiple-pulse excitation method is two-dimensional coherent spectroscopy (2DCS).
Initially introduced in the field of quantum chemistry \cite{Aue1976, Jeener1979, Tokmakoff2000, Mukamel2004}, this technique has been increasingly applied in physics \cite{Liu2025}---both experimentally \cite{Woerner2013, Somma2016, Singh2016a, Suzuki2016, Lu2017d, Liu2019q, Mahmood2021, Maehrlein2021, Mornhinweg2021, Blank2023, Zhang2024a, Zhang2024, Liu2024, Huang2024, Katsumi2024, Fujimoto2024, Barbalas2025, Cheng2025, Taherian2025} and theoretically \cite{Wan2019a, Liu2020, Choi2020c, Nandkishore2021, Phuc2021, Li2021, Li2023, Cai2022, Hu2022, Hart2023, Gao2023, Sim2023, Sim2023b, Wang2023, Negahdari2023, GomezSalvador2024, Hu2024, Watanabe2024, Watanabe2025, Potts2024, Potts2024b, Zhang2024b, Qiang2024, Puviani2024, Mootz2024, Valmispild2024, Chen2025, Negahdari2025, Salvador2025, Srivastava2025, DeWit2025, Sharma2025}---to study various systems.
In 2DCS, two short pulses are applied, and their time delay is swept to obtain the \textit{full} second-order response functions and the \textit{partial} third-order response functions.
Here, ``full'' indicates that the second-order response functions are resolved across the entire two-dimensional frequency space, while ``partial'' means that the obtained third-order response functions depend only on two frequency variables as they are integrated along one direction in three-dimensional frequency space [for details, see Supplemental Material (SM) \footnote{See Supplemental Material for the derivation of Eqs.~\eqref{eq:chi_nth}, \eqref{eq:derivative_n}, \eqref{eq:derivative_n-1}, and \eqref{eq:response_2nd_integral}; the relationship to two-dimensional coherent spectroscopy; the expressions for the response functions in quantum systems and for nonlinear optical responses; and the Hartree--Fock analysis of the Rice--Mele--Hubbard model.}].
By generalizing 2DCS to multidimensional coherent spectroscopy with multiple pulses, it is, in principle, possible to obtain the full $n$th-order response functions.
However, sweeping the time delays between the pulses significantly increases the computational cost, which has so far prevented numerical studies from obtaining the full higher-order response functions.

\begin{figure}[b]\centering
\includegraphics[scale=1]{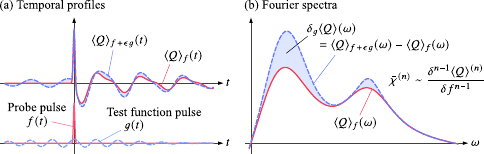}
\caption{(a)~Schematic of the temporal profiles of a physical quantity excited by pulsed fields $f(t)$ and $f(t) + \epsilon g(t)$.
(b)~The functional derivative is evaluated from the variation $\langle Q \rangle_{f+\epsilon g}(\omega) - \langle Q \rangle_{f}(\omega)$.
}
\label{fig:concept}
\end{figure}

In this Letter, we propose a distinct approach based on the functional derivative to obtain nonlinear response functions.
This framework enables the direct calculation of the \textit{full} $n$th-order response functions by simulating the time evolution of a system subjected to a temporally localized probe pulse and frequency-localized test function pulses (see Fig.~\ref{fig:concept}).
We demonstrate the validity of this method by computing the second- and third-order optical responses of the Rice--Mele model, showing quantitative agreement with perturbative expressions.
Furthermore, we illustrate its applicability to many-body systems using a tensor network method.
Since our formalism is general and provides nonlinear dynamical response functions solely by calculating the time evolution of physical quantities, without requiring the computation of multipoint correlation functions, it is compatible with any method capable of simulating real-time dynamics.

\pdfbookmark[1]{Formulation}{formulation}\textit{Formulation}---%
Consider a system whose Hamiltonian is written as $\mathcal{H}(t) = \mathcal{H}_0 + \mathcal{V}(t)$ at time $t$, where $\mathcal{H}_0$ is the unperturbed part and $\mathcal{V}(t)$ describes the coupling (which can be linear or nonlinear) between a physical quantity $Q_\alpha$ and time-dependent external fields $\{f_{\beta}(t)\}$ with $\alpha$ and $\beta$ labeling their respective components.
In general, the frequency response of $Q_\alpha$ to the external fields is expressed as $\langle Q_{\alpha} \rangle(\omega) = \sum_{n=0}^{\infty} \langle Q_{\alpha} \rangle^{(n)}(\omega)$ with 
\begin{align}
&\langle Q_{\alpha} \rangle^{(n)} (\omega) \notag \\
&= \sum_{\{\beta_i\}} \int_{-\infty}^{\infty} \frac{\mathrm{d}\omega_1 \cdots \mathrm{d}\omega_n}{(2\pi)^{n-1}} \delta(\omega_1 + \cdots + \omega_n - \omega) \notag \\
&\quad \times \chi_{\alpha\beta_1 \cdots \beta_n}^{(n)}(\omega_1, \dots, \omega_n) f_{\beta_1}(\omega_1) \cdots f_{\beta_n}(\omega_n)
\label{eq:chi_nth}
\end{align}
being the $n$th-order response \cite{Note1}.
Here, $\delta$ is the delta function, and $\chi^{(n)}$ denotes the $n$th-order retarded response function, which is given by the $(n-1)$th functional derivative of $\langle Q_{\alpha} \rangle^{(n)}(\omega)$ with respect to the external fields as follows:
\begin{align}
&\bar{\chi}_{\alpha\alpha_1 \cdots \alpha_{n}}^{(n)}(\omega_1, \dots, \omega_{n}) \notag \\
&= (2\pi)^{n-1} \frac{\partial}{\partial f_{\alpha_n}(\omega_n)} \frac{\delta^{n-1} \langle Q_\alpha \rangle^{(n)}(\omega)}{\delta f_{\alpha_{n-1}}(\omega_{n-1}) \cdots \delta f_{\alpha_1}(\omega_1)}, \label{eq:derivative_n}
\end{align}
with $\omega_n = \omega-\sum_{i=1}^{n-1}\omega_i$.
Here, $\bar{\chi}^{(n)}$ is the symmetric response function defined by
\begin{align}
&\bar{\chi}_{\alpha\alpha_1 \cdots \alpha_{n}}^{(n)}(\omega_1, \dots, \omega_{n}) \notag \\
&= \sum_{\sigma \in \mathfrak{S}_n} \chi_{\alpha\alpha_{\sigma(1)} \cdots \alpha_{\sigma(n)}}^{(n)}(\omega_{\sigma(1)}, \dots, \omega_{\sigma(n)}),
\label{eq:symmetric}
\end{align}
where $\mathfrak{S}_n$ is the symmetric group of degree $n$, reflecting the trivial permutation symmetry in Eq.~\eqref{eq:chi_nth}.
Hereafter, for simplicity, we consider the response of a single quantity $Q$ to a scalar field $f(t)$ and omit the index $\alpha$.
Then, Eq.~\eqref{eq:derivative_n} simplifies to \cite{Note1}
\begin{align}
\bar{\chi}^{(n)}(\omega_1, \dots, \omega_{n}) = \frac{(2\pi)^{n-1}}{f(\omega_n)} \frac{\delta^{n-1} \langle Q \rangle^{(n)}(\omega)}{\delta f(\omega_{n-1}) \cdots \delta f(\omega_1)}.
\label{eq:derivative_n-1}
\end{align}
Extending the following discussion to the general case [Eq.~\eqref{eq:derivative_n}] is straightforward.

To numerically evaluate the functional derivative in Eq.~\eqref{eq:derivative_n} or \eqref{eq:derivative_n-1}, it is necessary to calculate the time evolution of $\langle Q \rangle(t)$, which is triggered by an external field.
Here, we consider a Gaussian probe pulse,
\begin{align}
f(t) = \frac{F_0}{2\pi} \mathrm{e}^{-t^2/(2\tau_f^2)}, \quad
f(\omega) = \frac{F_0 \tau_f}{\sqrt{2\pi}} \mathrm{e}^{-\omega^2 \tau_f^2 /2},
\label{eq:field_f}
\end{align}
where $F_0$ and $\tau_f$ are the amplitude and the temporal width of the probe pulse, respectively.
The width $\tau_f$ must be sufficiently small [see also Fig.~\ref{fig:concept}(a)], so $f(\omega)$ remains nearly constant over the relevant energy range.
Once the $m$th-order response $\langle Q \rangle^{(m)}(\omega) \propto F_0^{m}$ is obtained, the $n$th-order response ($n > m$) can be determined as
\begin{align}
\langle Q \rangle^{(n)}(\omega)
= \langle Q \rangle(\omega) - \sum_{m=0}^{n-1} \langle Q \rangle^{(m)}(\omega)
= \mathcal{O}(f^n), \label{eq:Q_nth}
\end{align}
and
\begin{align}
\left. \langle Q \rangle^{(m)}(\omega) \right\vert_{F_0=F_0^{(n)}} = \left[ \frac{F_0^{(n)}}{F_0^{(n-1)}} \right]^{m} \left. \langle Q \rangle^{(m)}(\omega) \right\vert_{F_0=F_0^{(n-1)}}.
\label{eq:Q_diff}
\end{align}
Here, $F_0^{(n)}$ denotes the amplitude of $f$ in Eq.~\eqref{eq:Q_nth} and should be higher than $F_0^{(n-1)}$.
The first-order response function is given by $\chi^{(1)}(\omega) = \langle Q \rangle^{(1)}(\omega)/f(\omega)$ for a sufficiently small amplitude, $F_0 = F_0^{(1)}$.

For the second-order response function, the first functional derivative is required.
Let $g$ be a test function, $\epsilon$ be a small real number, and $\langle Q_{\alpha} \rangle^{(n)}$ be written as $\langle Q_{\alpha} \rangle_{f}^{(n)}$ to emphasize its dependence on $f$ as a functional.
The first variation of $\langle Q \rangle_{f}^{(2)}(\omega)$ is given by
\begin{align}
&\delta_{g} \langle Q \rangle^{(2)}(\omega)
\equiv \langle Q \rangle^{(2)}_{f+\epsilon g}(\omega) - \langle Q \rangle^{(2)}_{f}(\omega) \label{eq:response_2nd_def} \\
&= \epsilon \int_{-\infty}^{\infty} \mathrm{d}\omega_1\, g(\omega_1) \frac{\bar{\chi}^{(2)}(\omega_1,\omega-\omega_1) f(\omega-\omega_1)}{2\pi} + \mathcal{O}(\epsilon^2), \label{eq:response_2nd_integral}
\end{align}
as illustrated in Fig.~\ref{fig:concept}(b).
Here, the integrand, excluding $g(\omega_1)$, defines the functional derivative $\delta \langle Q \rangle^{(2)}(\omega)/\delta f(\omega_1)$.
Note that $\chi^{(2)}$ in Eq.~\eqref{eq:chi_nth} appears here in its symmetric form, $\bar{\chi}^{(2)}$ \cite{Note1}.
Since $g$ is arbitrary, this integral can be approximated by introducing test functions with sharp peaks in the frequency domain.
Specifically, we employ two types of Gaussian pulses:
\begin{align}
g_{\text{cos}_i}(t) &= \frac{F_0}{2\pi} \mathrm{e}^{-t^2/(2\tau_g^2)} \cos(\omega_i t), \label{eq:field_gcos} \\
g_{\text{sin}_i}(t) &= \frac{F_0}{2\pi} \mathrm{e}^{-t^2/(2\tau_g^2)} \sin(\omega_i t), \label{eq:field_gsin}
\end{align}
where $F_0$ is the same amplitude as in Eq.~\eqref{eq:field_f} and $\tau_{g}$ determines the frequency resolution.
The Fourier transform of $g_{\text{cos}_i}(t)$ is
\begin{align}
g_{\text{cos}_i}(\omega) &= \frac{F_0 \tau_g}{2\sqrt{2\pi}} \left[ \mathrm{e}^{-(\omega-\omega_i)^2 \tau_g^2/2} + \mathrm{e}^{-(\omega+\omega_i)^2 \tau_g^2/2} \right],
\end{align}
which reduces to $(F_0/2) [\delta(\omega-\omega_i) + \delta(\omega+\omega_i)]$ in the limit of $\tau_g \to \infty$.
Thus, for sufficiently large $\tau_g$, the integral in Eq.~\eqref{eq:response_2nd_integral} is evaluated as
\begin{align}
\frac{\delta_{g} \langle Q \rangle^{(2)}(\omega)}{\epsilon F_0^{(2)}}
&\approx \frac{1}{4\pi} \bar{\chi}^{(2)}(\omega_1,\omega-\omega_1) f(\omega-\omega_1) \notag \\
&\quad + \frac{1}{4\pi} \bar{\chi}^{(2)}(\omega_1,\omega+\omega_1) f(\omega+\omega_1)
\end{align}
for $g=g_{\text{cos}_1}$ and similarly for $g=g_{\text{sin}_1}$.
Finally, we obtain the formula for the second-order response as
\begin{align}
&\bar{\chi}^{(2)}(\pm\omega_1,\omega\mp \omega_1) \notag \\
&= \frac{2\pi}{f(\omega\mp \omega_1)} \left[ \frac{\delta_{\text{cos}_1} \langle Q \rangle^{(2)}(\omega)}{\epsilon F_0^{(2)}} \mp \mathrm{i} \frac{\delta_{\text{sin}_1} \langle Q \rangle^{(2)}(\omega)}{\epsilon F_0^{(2)}} \right].
\label{eq:chi2}
\end{align}
Here, $\delta_{\text{cos}_1} \langle Q \rangle^{(2)}(\omega)$ and $\delta_{\text{sin}_1} \langle Q \rangle^{(2)}(\omega)$ are the variations defined in Eq.~\eqref{eq:response_2nd_def} where $g$ is replaced with $g_{\text{cos}_1}$ and $g_{\text{sin}_1}$, respectively.
These quantities can be numerically obtained from $\langle Q \rangle^{(2)}(\omega)$ under three types of external fields: $f(t) + \epsilon g_{\text{cos}_1}(t)$, $f(t) + \epsilon g_{\text{sin}_1}(t)$, and $f(t)$.

The third-order response requires the second functional derivative.
By introducing two test functions $g_1$ and $g_2$, we find that the second variation of $\langle Q \rangle^{(3)}(\omega)$ is given by
\begin{align}
&\delta_{g_2} \delta_{g_1} \langle Q \rangle^{(3)}(\omega) \notag \\
&\equiv \langle Q \rangle^{(3)}_{f+\epsilon g_1 + \epsilon g_2}
- \langle Q \rangle^{(3)}_{f+\epsilon g_2}
- \langle Q \rangle^{(3)}_{f+ \epsilon g_1}
+ \langle Q \rangle^{(3)}_{f} \label{eq:response_3rd_def} \\
&= \epsilon^2 \int_{-\infty}^{\infty} \mathrm{d}\omega_1 \mathrm{d}\omega_2\, g_1(\omega_1) g_2(\omega_2) \notag \\
&\quad \times \frac{\bar{\chi}^{(3)}(\omega_1,\omega_2,\omega-\omega_1-\omega_2) f(\omega-\omega_1-\omega_2)}{(2\pi)^2} + \mathcal{O}(\epsilon^3). \label{eq:response_3rd_integral}
\end{align}
To evaluate the integral, we adopt the test functions in Eqs.~\eqref{eq:field_gcos} and \eqref{eq:field_gsin} and assume that $\tau_g$ is sufficiently large.
Using the same approach as for $\bar{\chi}^{(2)}$, we obtain the following formula:
\begin{align}
&\bar{\chi}^{(3)}(\omega_1,\omega_2,\omega_3) \notag \\
&= \frac{(2\pi)^2}{f(\omega_3) [\epsilon F_0^{(3)}]^2} \bigl[ \delta_{\text{cos}_2} \delta_{\text{cos}_1} \langle Q \rangle^{(3)}(\omega) - \delta_{\text{sin}_2} \delta_{\text{sin}_1} \langle Q \rangle^{(3)}(\omega) \notag \\
&\quad - \mathrm{i} \delta_{\text{cos}_2} \delta_{\text{sin}_1} \langle Q \rangle^{(3)}(\omega) - \mathrm{i} \delta_{\text{sin}_2} \delta_{\text{cos}_1} \langle Q \rangle^{(3)}(\omega) \bigr],
\label{eq:chi3}
\end{align}
where $\omega_3 = \omega-\omega_1-\omega_2$.
Thus, $\bar{\chi}^{(3)}$ can be calculated from $\langle Q \rangle^{(3)}(\omega)$ under seven ($= 2^2 + 2 + 1$) types of external fields.
By successively applying a similar procedure, the expressions for higher-order response functions can also be derived.
However, ensuring numerical precision---particularly when subtracting lower-order responses in Eq.~\eqref{eq:Q_nth}---will become increasingly challenging.

\pdfbookmark[1]{Proof-of-principle demonstration}{demonstration}\textit{Proof-of-principle demonstration}---%
To validate the present framework, we calculate the nonlinear optical responses of an electron system described by the spinless Rice--Mele model.
The Hamiltonian is given by $\mathcal{H}_0 = \sum_{k} \varPsi_k^\dagger H_k \varPsi_k$ with $H_k = -h_{x} \cos(k/2) \tau_x - h_{y} \sin(k/2) \tau_y - h_z \tau_z$, where $\tau$ denotes the Pauli matrix and $\varPsi_k^\dagger = (c_{k\uparrow}^\dagger, c_{k\downarrow}^\dagger)$ represents the creation operators for electrons with momentum $k$ and pseudospin up and down.
We consider the half-filling case, where the energy gap between the conduction and valence bands is $E_{\mathrm{g}} = 2(h_y^2 + h_z^2)^{1/2}$ for $|h_y| \leq |h_x|$.
When both $h_y/h_x$ and $h_z/h_x$ are finite and nonzero, inversion symmetry is broken, and even-order responses are allowed.
Throughout this Letter, we set $h_x = 1$ as the unit of energy and take the Dirac constant $\hbar$, the lattice constant, and the electron charge to be unity.

The optical vector potential $A(t)$ is introduced through Peierls substitution: $H_{k} \to H_{k-A(t)}$, and the electric field is given by $E(t) = -\partial_t A(t)$.
In this velocity gauge, the coupling between the vector potential and the electric current is nonlinear.
The time evolution of the system is governed by the von Neumann equation: $\partial_t \rho_k(t) = -\mathrm{i} [H_{k-A(t)}, \rho_k(t)] - \varGamma [\rho_{k}(t) - \rho_{k,0}(t)]$, where $\rho_{k,ss'} = \langle c_{ks'}^\dagger c_{ks} \rangle$ represents the one-body density matrix, and $\rho_{k,0}$ denotes the density matrix in the ground state of $H_{k-A(t)}$.
We use the relaxation time approximation (RTA) to quantitatively validate the present framework against perturbative formulas;
$\varGamma\ (=0.01)$ corresponds to the relaxation rate of electrons.
The electric current density is calculated as $\langle J \rangle(t) = N^{-1} \sum_k \Tr[\rho_k \partial_k H_{k-A(t)}]$, where $N\ (=800)$ is the number of $k$ points.
For the time evolution, we employ the fourth-order Runge--Kutta method (with a time step of $0.01$) and double-double arithmetic implemented in the Julia package \texttt{DoubleFloats.jl}, which efficiently improves numerical precision.

\begin{figure}[t]\centering
\includegraphics[scale=1]{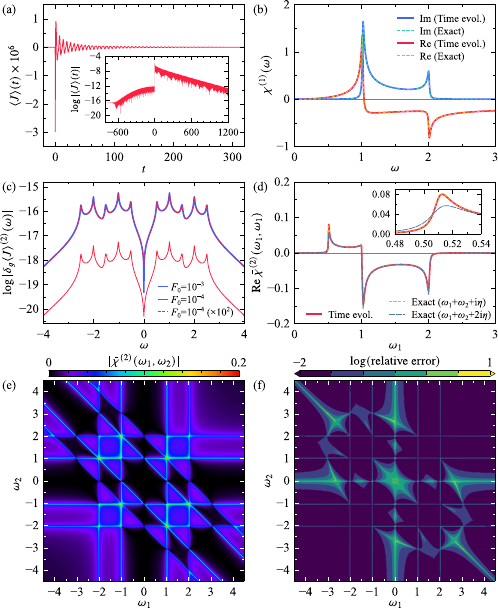}
\caption{(a)~Temporal profile of the electric current density $\langle J \rangle(t)$.
The inset shows $\vert \langle J \rangle(t) \vert$ on a logarithmic scale.
(b)~First-order response function obtained from time evolution (solid lines) in comparison with the exact result (dashed lines).
(c)~First variation of the current [Eq.~\eqref{eq:response_2nd_def}] for $F_0 = 10^{-4}$ (red) and $F_0 = 10^{-3}$ (blue) with $g= g_{\text{cos}_1}$ and $\omega_1 = 0.5$.
(d)~Second-order response function obtained from the time evolution (red solid line) and the perturbative expression (dashed lines).
The inset is an enlargement for $\omega_1 \in [0.48, 0.54]$.
[(e) and (f)]~Contour maps of (e) $\bar{\chi}^{(2)}(\omega_1, \omega_2)$ and (f) its logarithmic relative error.}
\label{fig:second}
\end{figure}

First, we present in Fig.~\ref{fig:second}(a) the temporal profile of the current density when the vector potential $A(t) = f(t) + \epsilon g_{\text{cos}_1}(t)$ is applied.
The system is in the broken-inversion-symmetry ferroelectric phase with $h_y = 0.5$ and $h_z = 0.1$, and the pulse parameters are set to $F_0 = 10^{-4}$, $\epsilon = 10^{-6}$, $\tau_{f} = 0.1$, $\tau_{g} = \sqrt{2}/\varGamma$, and $\omega_1 = 0.5$.
Since the test function $g_{\text{cos}_1}(t)$ has a broad envelope in the time domain, the current begins to rise around $t \sim -600$ and is then excited by the delta-function-like pulse $f(t)$ at $t=0$.
After this excitation, the current decays exponentially, governed by the relaxation rate $\varGamma = 0.01$.
By performing the Fourier transformation of these temporal profiles, the variation in the current, $\delta \langle J \rangle(\omega)$, can be calculated.

Before discussing nonlinear responses, we show the first-order response in Fig.~\ref{fig:second}(b) as a reference.
This is calculated from $\chi^{(1)}(\omega) = \langle J \rangle^{(1)}(\omega)/A(\omega)$, with $A(t) = f(t)$ and $F_0 = 10^{-6}$.
The optical gap is $E_{\mathrm{g}} = 1.02$, and a continuum is observed in the range $E_{\mathrm{g}} \lesssim \omega \lesssim 2$, corresponding to the energy bandwidth.
The result obtained from the time evolution shows exact agreement with those from the perturbative expression [Eq.~(S52) in SM \cite{Note1}] where the imaginary part of the frequency, called the broadening factor, is set to $\eta = \varGamma = 0.01$.

To obtain the second-order response function from Eq.~\eqref{eq:chi2}, we need the first variation defined in Eq.~\eqref{eq:response_2nd_def}.
Figure~\ref{fig:second}(c) shows $\delta_{\text{cos}_1} \langle J \rangle^{(2)}(\omega)$ for $F_0 = 10^{-4}$ and $10^{-3}$, with $\epsilon = 10^{-6}$ and $\omega_1 = 0.5$.
As expected, the result for $F_0 = 10^{-3}$ (blue line) agrees well with the result for $F_0 = 10^{-4}$ multiplied by $10^2$ (red dashed line), indicating that the variation is indeed proportional to $F_0^2$.
In practical calculations, it is necessary to confirm such a scaling law with respect to $F_0$ or to verify that the obtained response function is independent of $F_0$ and $\epsilon$.

Equation~\eqref{eq:chi2} for the second-order response function depends on the frequency of the test function, $\omega_1$.
By sweeping $\omega_1$, we can obtain the full two-dimensional response function $\bar{\chi}^{(2)}(\omega_1, \omega_2)$.
From this, specific responses such as the second harmonic response $\bar{\chi}^{(2)}(\omega_1, \omega_1)$ or the rectification response $\bar{\chi}^{(2)}(\omega_1, -\omega_1)$ can be extracted.
In Fig.~\ref{fig:second}(d), the second harmonic response is shown, which agrees closely with those obtained from the perturbative expression.

It has been recognized that subtle issues arise regarding how to introduce a small imaginary part to the frequency when calculating nonlinear response functions based on perturbative expressions \cite{Passos2018}.
Specifically, in the energy denominator, terms such as $\omega_1 + \omega_2$ appear [see Eq.~(S41) in SM \cite{Note1}], and one reasonable approach is to replace them with $\omega_1 + \omega_2 + 2\mathrm{i}\eta$, while another possible approach is to use $\omega_1 + \omega_2 + \mathrm{i}\eta$.
In Fig.~\ref{fig:second}(d), the former is plotted as the blue dashed line, while the latter is shown as the orange dashed line.
Overall, the two approaches yield similar results.
However, notable differences appear around $\omega_1 \approx 0.51$, corresponding to the two-photon absorption peak near the band edge, where $\omega_1 + \omega_2 \approx E_{\mathrm{g}}$.
In this region, the results calculated with $\omega_1 + \omega_2 + \mathrm{i}\eta$ show better agreement, at least within the RTA.

Figure~\ref{fig:second}(e) shows a contour map of the second-order response function, $\bar{\chi}^{(2)}(\omega_1, \omega_2)$.
Reflecting the Van Hove singularities at $\omega \approx 1$ and $2$ shown in Fig.~\ref{fig:second}(b), strong intensities are observed along the lines $\omega_1, \omega_2 \approx \pm 1, 2$ and $\omega_1 + \omega_2 \approx \pm 1, 2$, while continua appear in the regions enclosed by these lines.
The results in Fig.~\ref{fig:second}(d) correspond to a cut along the line $\omega_1 = \omega_2$.
Figure~\ref{fig:second}(f) displays the relative error with respect to the result obtained from the perturbative expression [Eq.~(S55) in SM \cite{Note1}] where $\omega_1^+ + \omega_2^+$ is replaced with $\omega_1 + \omega_2 + \mathrm{i}\eta$.
In most regions, the relative error is below $10^{-1.5} = 3.2\%$, although it increases to $10^{-1} = 10\%$ or at most $10^{-0.5} = 32\%$ near the Van Hove singularities.
Outside these regions, along $\omega_1 = 0$, $\omega_2 = 0$, and $\omega_2 = -\omega_1$, the relative error increases simply because $\bar{\chi}^{(2)}$ approaches zero, while the absolute error remains negligible.
These results demonstrate that the present method accurately reconstructs the second-order response functions in the whole two-dimensional frequency space, with small discrepancies primarily limited to regions near the singularities.

\begin{figure}[t]\centering
\includegraphics[scale=1]{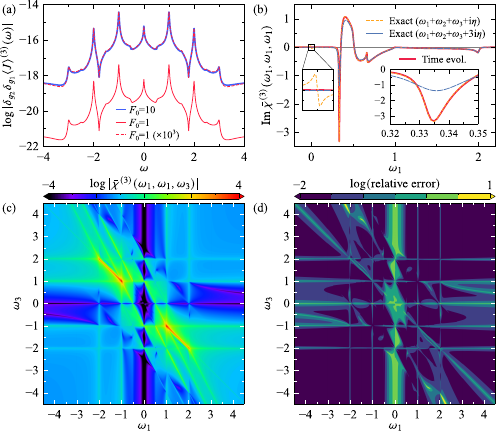}
\caption{(a)~Second variation of the current [Eq.~\eqref{eq:response_3rd_def}] for $F_0 = 1$ (red) and $F_0 = 10$ (blue) with $g_1 = g_{\text{cos}_1}$, $g_2 = g_{\text{cos}_2}$, and $\omega_1 = \omega_2 = 0.5$.
(b)~Third-order response function obtained from time evolution (red solid line) in comparison with the perturbative expression (dashed lines).
The insets show enlargements for $\omega_1 \in [-0.04, 0.04]$ and $[0.32, 0.35]$.
[(c) and (d)]~Contour maps of (c) $\bar{\chi}^{(3)}(\omega_1, \omega_1, \omega_3)$ and (d) its logarithmic relative error.}
\label{fig:third}
\end{figure}

Next, we discuss the third-order response.
The model parameters are set to $h_y = 0.5$ and $h_z = 0$, so the second-order response vanishes because of the presence of inversion symmetry.
To evaluate the third-order response in Eq.~\eqref{eq:chi3}, the second variations defined in Eq.~\eqref{eq:response_3rd_def} are required.
Figure~\ref{fig:third}(a) shows $\delta_{\text{cos}_2} \delta_{\text{cos}_1} \langle J \rangle^{(3)}(\omega)$, calculated with $\epsilon = 10^{-8}$ and $\omega_1 = \omega_2 = 0.5$.
The result obtained for $F_0 = 1$ multiplied by $10^3$ (red dashed line) agrees well with that for $F_0 = 10$ (blue line), confirming that the variation scales as $F_0^3$, as expected for the third-order responses.

Although the full third-order response function can be obtained by sweeping the frequencies of the two test functions, $\omega_1$ and $\omega_2$, we focus here on the case of $\omega_1 = \omega_2$.
Figure~\ref{fig:third}(b) shows the imaginary part of the third harmonic response, $\bar{\chi}^{(3)}(\omega_1, \omega_1, \omega_1)$.
The most prominent peak occurs at $\omega_1 = 0.335 \approx E_{\mathrm{g}}/3$, which corresponds to the three-photon absorption peak at the band edge.
In this case as well, good agreement is observed with the perturbative expression [Eq.~(S57) in SM \cite{Note1}] where the same broadening factor $\mathrm{i}\eta$ is added to, e.g., $\omega_1$, $\omega_1 + \omega_2$, and $\omega_1 + \omega_2 + \omega_3$ (orange dashed line).
However, around $\omega_1 = 0$, better agreement is achieved when $2\mathrm{i}\eta$ and $3\mathrm{i}\eta$ are added to $\omega_1 + \omega_2$ and $\omega_1 + \omega_2 + \omega_3$, respectively (blue dashed line).

In Fig.~\ref{fig:third}(c), we present the contour map of $\bar{\chi}^{(3)}(\omega_1, \omega_1, \omega_3)$ on a logarithmic scale.
Pronounced intensities are observed along lines corresponding to the Van Hove singularities.
In particular, the region along $\omega_3 = -\omega_1$ for $1 \lesssim \vert \omega_1 \vert \lesssim 2$ exhibits significantly large peaks associated with multiple resonances.
Figure~\ref{fig:third}(d) shows the relative error with respect to the perturbative expression that adopts the same broadening factor $\mathrm{i}\eta$.
Similar to the second-order response, the error increases in regions where the Van Hove singularities appear or where $\bar{\chi}^{(3)}$ approaches zero.
Nevertheless, the error remains below a few percent overall, demonstrating that the present framework can accurately obtain the third-order response functions.

\pdfbookmark[1]{Application to many-body systems}{application}\textit{Application to many-body systems}---%
As a demonstration of the applicability, we calculate the second-order responses in the spinful Rice--Mele--Hubbard model \cite{Nakagawa2018a}, by employing the infinite time-evolving block decimation (iTEBD) algorithm \cite{Vidal2007}.
This is an unbiased, nonperturbative method that can compute both the ground state and the time evolution of one-dimensional many-body systems in the thermodynamic limit.
See End Matter for details on the model and computations.

\begin{figure}[t]\centering
\includegraphics[scale=1]{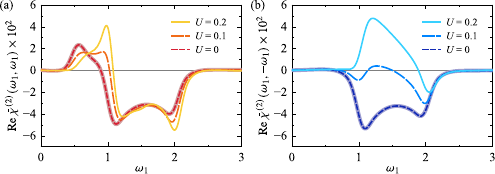}
\caption{Second-order response functions: (a) $\bar{\chi}^{(2)}(\omega_1, \omega_1)$ and (b) $\bar{\chi}^{(2)}(\omega_1, -\omega_1)$ for $U = 0, 0.1, 0.2$, calculated using iTEBD with a bond dimension of $D = 1200$.
The bold gray curves represent results obtained by solving the von Neumann equation with $U = 0$ and $\varGamma = 0$.
}
\label{fig:tebd}
\end{figure}

Figure~\ref{fig:tebd} shows the second harmonic and rectification responses for different values of the interaction strength $U$, with $h_y = 0.5$ and $h_z = 0.1$.
When $U=0$, the iTEBD results agree with those in the Rice--Mele model obtained from the von Neumann equation.
Introducing $U$ modifies the spectral shapes, and for $U = 0.2\ (> h_z)$, a sign reversal is observed in $\chi^{(2)}(\omega_1,-\omega_1)$, but not in $\chi^{(2)}(\omega_1, \omega_1)$.
Considering that the effect of $U$ is renormalized into $h_z$ in the Hartree--Fock approximation \cite{Note1}, this selective sign reversal reflects a many-body effect beyond one-body physics.
While such nonlinear responses in many-body systems require further investigation, these results demonstrate the versatility and computational feasibility of our approach.

\pdfbookmark[1]{Discussion}{discussion}\textit{Discussion}---%
In the present framework, if convergence is achieved with respect to parameters such as $F_0$, $\epsilon$, and $\tau_g$, the resulting response function can be considered accurate within the approximation of the adopted time-evolution method.
When both time-evolution calculations and perturbative calculations are feasible, it becomes possible to compare and evaluate the validity of the former's approximation or the latter's self-energy and vertex corrections, thereby providing deeper insights into dissipation and scattering processes.
In the examples shown in Figs.~\ref{fig:second} and \ref{fig:third}, quantitative agreement is observed with the results obtained by introducing a common broadening factor $\mathrm{i} \eta$ into the perturbative expressions.
This agreement may suggest that, in the RTA, higher moments of the current operator decay with the same relaxation rate $\varGamma$ as the first moment $\langle J \rangle(t)$.
Investigating dissipation effects on nonlinear responses using more advanced methods beyond the RTA represents an interesting direction for future research.

Additionally, since a single probe pulse $f(t)$ is applied in this method, transient dynamics can be investigated by exciting a system with a pump pulse and varying the time delay of the probe pulse.
This approach enables the definition of the transient nonlinear response functions as a natural extension of the pump-probe-based transient linear response \cite{Shao2016}.
This point will be examined in a follow-up study.

\begin{acknowledgments}
\textit{Acknowledgments}---%
This work was supported by JSPS KAKENHI Grants No.\ JP23K13052, No.\ JP23K25805, and No.\ JP24K00563.
The numerical calculations were performed using the facilities of the Supercomputer Center, the Institute for Solid State Physics, the University of Tokyo.
\end{acknowledgments}

\bibliography{reference,reference_unpublished}

\onecolumngrid
\section*{End Matter}
\setcounter{equation}{0}
\renewcommand{\theequation}{A\arabic{equation}}
\twocolumngrid
\pdfbookmark[2]{Appendix}{appendix}\textit{Appendix}---%
The Hamiltonian of the spinful Rice--Mele--Hubbard model on a one-dimensional chain is given by
\begin{align}
\mathcal{H}_{\mathrm{RMH}} &= - \sum_{js} \frac{h_x - (-1)^{j} h_y}{2} \left( \mathrm{e}^{\mathrm{i} A(t)/2} c_{j+1,s}^\dagger c_{js} + \mathrm{H.c.} \right) \notag \\
&\quad - h_z \sum_{js} (-1)^j c_{js}^\dagger c_{js} \notag \\
&\quad + U \sum_{j} \left( c_{j\uparrow}^\dagger c_{j\uparrow} - \frac{1}{2} \right) \left( c_{j\downarrow}^\dagger c_{j\downarrow} - \frac{1}{2} \right),
\label{eq:rmh}
\end{align}
where $c_{js}^\dagger$ is the creation operator for an electron with spin $s$ at site $j$ and $U$ represents the Hubbard interaction strength.
The parameters $(h_x, h_y, h_z)$ are the same as those in the spinless Rice--Mele model introduced in the main text.
The vector potential is denoted by $A(t)$, and the nearest-neighbor site distance is set to $1/2$.
The electric current is defined as $\langle J \rangle(t) = - N^{-1} \langle \partial \mathcal{H}_{\mathrm{RMH}} / \partial A \rangle$, which coincides with that of the spinless Rice--Mele model when $U = 0$.

The many-body state can be approximated by a matrix product state, whose bond dimension $D$ serves as a parameter controlling the accuracy.
The ground state and its time evolution are obtained using the iTEBD algorithm, implemented with \texttt{ITensor} \cite{Fishman2022}.
We employ the fourth-order Suzuki--Trotter decomposition with a time step of $0.0625$ and exploit the global $\mathrm{U}(1) \otimes \mathrm{U}(1)$ symmetry.
Since iTEBD does not incorporate relaxation effects, we apply a Gaussian window function, $w(t) = \exp[-t^2/(2\tau_w^2)]$, to the temporal profiles of $\delta \langle J \rangle(t)$.
If we instead use an exponential window function, e.g., $\exp(-\varGamma |t|)$, we can directly compare the results with the perturbative expression; however, using a Gaussian window allows for faster convergence even in short time calculations.

\begin{figure}[t]\centering
\includegraphics[scale=1]{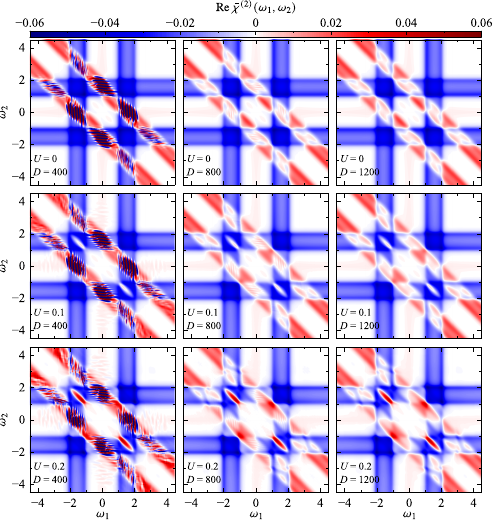}
\caption{Contour maps of $\bar{\chi}^{(2)}(\omega_1, \omega_2)$ for $U = 0, 0.1, 0.2$ (top to bottom) and $D = 400, 800, 1200$ (left to right).
}
\label{fig:u-d}
\end{figure}

Figure~\ref{fig:u-d} shows $\re \bar{\chi}^{(2)}(\omega_1, \omega_2)$ for different values of $U$ and $D$.
The other parameters are set to $h_y = 0.5$, $h_z = 0.1$, $F_0^{(2)} = 10F_0^{(1)} = 0.1$, $\epsilon = 10^{-3}$, $\tau_f = 0.125$, $\tau_g = 12$, and $\tau_w = 6$, which are the same as those used in Fig.~\ref{fig:tebd}.
The convergence behavior with respect to $D$ depends on $\omega_1$ and $\omega_2$ but is largely unaffected by $U$.
In particular, the second harmonic response along $\omega_1 = \omega_2$ and the rectification response along $\omega_1 = -\omega_2$ exhibit good convergence even at $D = 800$.
However, around $(\omega_1, \omega_2) = (1.5, 0)$, where the positive and negative continua interfere destructively, convergence becomes slower.
This highlights the importance of examining the convergence with respect to $D$, as well as $F_0$ and $\epsilon$.

\end{document}


\title{\texorpdfstring{Supplemental Material for \\``Extracting Nonlinear Dynamical Response Functions from Time Evolution''}{Supplemental Material for “Extracting Nonlinear Dynamical Response Functions from Time Evolution”}}
\author{\mbox{Atsushi Ono}}
\affiliation{\mbox{Department of Physics, Tohoku University, Sendai 980-8578, Japan}}

\maketitle


\section{Derivation of equations in the main text}
\subsection{Equation (1)}
The perturbative response of a physical quantity $\langle Q_{\alpha} \rangle(t)$ to external fields $\{f_{\beta}(t)\}$ is generally written as
\begin{align}
\langle Q_\alpha \rangle(t) &= \sum_{n=0}^{\infty} \langle Q_\alpha \rangle^{(n)}(t), \quad
\langle Q_\alpha \rangle^{(n)}(t)
= \sum_{\{\beta_i\}} \int_{-\infty}^{\infty} \mathrm{d}t_1 \cdots \mathrm{d}t_n \, \chi_{\alpha\beta_1\cdots \beta_n}^{(n)}(t;t_1,\dots,t_n) f_{\beta_1}(t_1) \cdots f_{\beta_n}(t_n).
\label{eq:response_time}
\end{align}
Here, $\chi_{\alpha\beta_1\cdots \beta_n}^{(n)}(t;t_1,\dots,t_n)$ is the $n$th-order retarded response function, which is nonzero when $t\geq t_1 \geq t_2 \geq \cdots \geq t_n$ and vanishes otherwise because of causality.
In addition, when the system is in a stationary state, the response function is invariant under time translation:
\begin{align}
\chi_{\alpha\beta_1\cdots \beta_n}^{(n)}(t;t_1,\dots,t_n)
= \chi_{\alpha\beta_1\cdots \beta_n}^{(n)}(t+T;t_1+T,\dots,t_n+T)
\label{eq:stationarity}
\end{align}
for any $T$.
Thus, the response function can be written as
\begin{align}
\chi_{\alpha\beta_1\cdots \beta_n}^{(n)}(t;t_1,\dots,t_n)
= \chi_{\alpha\beta_1\cdots \beta_n}^{(n)}(0;t_1-t,\dots,t_n-t)
\equiv \chi_{\alpha\beta_1\cdots \beta_n}^{(n)}(\bar{t}_1,\dots,\bar{t}_n),
\end{align}
with $\bar{t}_i = t-t_i$.
The Fourier transformation of $\chi_{\alpha\beta_1\cdots \beta_n}^{(n)}(t; t_1,\dots,t_n)$ is defined by
\begin{align}
\chi_{\alpha\beta_1\cdots \beta_n}^{(n)}(\omega; \omega_1,\dots,\omega_n)
&= \int_{-\infty}^{\infty} \mathrm{d}t \mathrm{d}t_1 \cdots \mathrm{d}t_n\, \mathrm{e}^{\mathrm{i}\omega t+\mathrm{i}\omega_1 t_1+\cdots +\mathrm{i}\omega_n t_n} \chi_{\alpha\beta_1\cdots \beta_n}^{(n)}(t; t_1,\dots,t_n),
\end{align}
and for stationarity [Eq.~\eqref{eq:stationarity}], it reduces to
\begin{align}
\chi_{\alpha\beta_1\cdots \beta_n}^{(n)}(\omega; -\omega_1,\dots,-\omega_n)
&= \int_{-\infty}^{\infty} \mathrm{d}t\, \mathrm{e}^{\mathrm{i}(\omega-\omega_1-\cdots-\omega_n)t} \int_{-\infty}^{\infty} \mathrm{d}\bar{t}_1 \cdots \mathrm{d}\bar{t}_n\, \mathrm{e}^{\mathrm{i}\omega_1 \bar{t}_1 + \cdots + \mathrm{i}\omega_n \bar{t}_n} \chi_{\alpha\beta_1\cdots \beta_n}^{(n)}(\bar{t}_1,\dots,\bar{t}_n) \notag \\
&= 2\pi\delta(\omega_1+\cdots+\omega_n-\omega) \chi_{\alpha\beta_1\cdots \beta_n}^{(n)}(\omega_1,\dots,\omega_n),
\end{align}
where
\begin{align}
\chi_{\alpha\beta_1\cdots \beta_n}^{(n)}(\omega_1,\dots,\omega_n)
&= \int_{-\infty}^{\infty} \mathrm{d}\bar{t}_1 \cdots \mathrm{d}\bar{t}_n\, \mathrm{e}^{\mathrm{i}\omega_1 \bar{t}_1 + \cdots + \mathrm{i}\omega_n \bar{t}_n} \chi_{\alpha\beta_1\cdots \beta_n}^{(n)}(\bar{t}_1,\dots,\bar{t}_n).
\label{eq:def_chi_nth}
\end{align}
Given the above, we obtain the expression for the $n$th-order frequency response as follows:
\begin{align}
\langle Q_{\alpha} \rangle^{(n)}(\omega)
&= \int_{-\infty}^{\infty} \mathrm{d}t\, \mathrm{e}^{\mathrm{i}\omega t} \langle Q_{\alpha} \rangle^{(n)}(t) \notag \\
&= \sum_{\{\beta_i\}} \int_{-\infty}^{\infty} \mathrm{d}t\, \mathrm{e}^{\mathrm{i}\omega t} \int_{-\infty}^{\infty} \mathrm{d}t_1 \cdots \mathrm{d}t_n\, \chi_{\alpha\beta_1\cdots \beta_n}^{(n)}(t;t_1,\dots,t_n) f_{\beta_1}(t_1) \cdots f_{\beta_n}(t_n) \notag \\
&= \sum_{\{\beta_i\}} \int_{-\infty}^{\infty} \mathrm{d}t\, \mathrm{e}^{\mathrm{i}\omega t} \int_{-\infty}^{\infty} \mathrm{d}t_1 \cdots \mathrm{d}t_n \int_{-\infty}^{\infty} \frac{\mathrm{d}\omega' \mathrm{d}\omega_1' \cdots \mathrm{d}\omega_n'}{(2\pi)^{n+1}}\, \mathrm{e}^{-\mathrm{i}\omega't-\mathrm{i}\omega_1' t_1-\cdots-\mathrm{i}\omega_n' t_n} \chi_{\alpha\beta_1\cdots \beta_n}^{(n)}(\omega'; \omega_1',\dots,\omega_n') \notag \\
&\quad \times \int_{-\infty}^{\infty} \frac{\mathrm{d}\omega_1 \cdots \mathrm{d}\omega_n}{(2\pi)^{n}}\, \mathrm{e}^{-\mathrm{i}\omega_1 t_1-\cdots-\mathrm{i}\omega_n t_n} f_{\beta_1}(\omega_1) \cdots f_{\beta_n}(\omega_n) \notag \\
&= \sum_{\{\beta_i\}} \int_{-\infty}^{\infty} \frac{\mathrm{d}\omega_1 \cdots \mathrm{d}\omega_n}{(2\pi)^{n}}\, \chi_{\alpha\beta_1\cdots \beta_n}^{(n)}(\omega; -\omega_1,\dots,-\omega_n) f_{\beta_1}(\omega_1) \cdots f_{\beta_n}(\omega_n) \notag \\
&= \sum_{\{\beta_i\}} \int_{-\infty}^{\infty} \frac{\mathrm{d}\omega_1 \cdots \mathrm{d}\omega_n}{(2\pi)^{n-1}}\, \delta(\omega_1+\cdots+\omega_n-\omega) \chi_{\alpha\beta_1\cdots \beta_n}^{(n)}(\omega_1,\dots,\omega_n) f_{\beta_1}(\omega_1) \cdots f_{\beta_n}(\omega_n).
\label{eq:sm_chi_nth}
\end{align}
This is Eq.~(1) in the main text.

\subsection{Equations (2) and (4)}
The $(n-1)$th functional derivative of $\langle Q_{\alpha} \rangle^{(n)}(\omega)$ in Eq.~\eqref{eq:sm_chi_nth} is
\begin{align}
\frac{\delta^{n-1} \langle Q_\alpha \rangle^{(n)}(\omega)}{\delta f_{\alpha_{n-1}}(\omega_{n-1}) \cdots \delta f_{\alpha_1}(\omega_1)}
= \sum_{\beta_n} \frac{\bar{\chi}_{\alpha\alpha_1 \cdots \alpha_{n-1} \beta_{n}}^{(n)}(\omega_1, \dots, \omega_{n}) f_{\beta_n}(\omega_n)}{(2\pi)^{n-1}},
\label{eq:n-1th}
\end{align}
where $\bar{\chi}^{(n)}$ is the $n$th-order symmetric response function defined in Eq.~(3).
For simplicity, we consider the second-order response and assume that the indices run over two Cartesian directions, namely, $\alpha, \beta \in \{x, y\}$.
In this case, we have
\begin{align}
\frac{\delta \langle Q_{x} \rangle^{(2)}(\omega)}{\delta f_{x}(\omega_1)}
= \sum_{\beta_2} \frac{\bar{\chi}_{x x \beta_{2}}^{(2)}(\omega_1, \omega_{2}) f_{\beta_2}(\omega_2)}{2\pi}
= \frac{1}{2\pi} \left[ \bar{\chi}_{x x x}^{(2)}(\omega_1, \omega_{2}) f_{x}(\omega_2) + \bar{\chi}_{x x y}^{(2)}(\omega_1, \omega_{2}) f_{y}(\omega_2) \right] .
\label{eq:1st_deriv}
\end{align}
Thus, differentiating with respect to $f_{\beta_2}(\omega_2)$ yields
\begin{align}
\frac{\partial}{\partial f_{x}(\omega_2)} \frac{\delta \langle Q_{x} \rangle^{(2)}(\omega)}{\delta f_{x}(\omega_1)}
= \frac{1}{2\pi} \bar{\chi}_{x x x}^{(2)}(\omega_1, \omega_{2}), \quad
\frac{\partial}{\partial f_{y}(\omega_2)} \frac{\delta \langle Q_{x} \rangle^{(2)}(\omega)}{\delta f_{x}(\omega_1)}
= \frac{1}{2\pi} \bar{\chi}_{x x y}^{(2)}(\omega_1, \omega_{2}).
\end{align}
By generalizing this result, we obtain Eq.~(2):
\begin{align}
\bar{\chi}_{\alpha\alpha_1 \cdots \alpha_{n}}^{(n)}(\omega_1, \dots, \omega_{n})
= (2\pi)^{n-1} \frac{\partial}{\partial f_{\alpha_n}(\omega_n)} \frac{\delta^{n-1} \langle Q_\alpha \rangle^{(n)}(\omega)}{\delta f_{\alpha_{n-1}}(\omega_{n-1}) \cdots \delta f_{\alpha_1}(\omega_1)}.
\label{eq:sm_derivative_n}
\end{align}
If we omit the indices, the summation in Eq.~\eqref{eq:n-1th} disappears, allowing us to obtain $\bar{\chi}^{(n)}(\omega_1, \dots, \omega_{n})$ by simply dividing Eq.~\eqref{eq:n-1th} by $f(\omega_n)$:
\begin{align}
\bar{\chi}^{(n)}(\omega_1, \dots, \omega_{n}) = \frac{(2\pi)^{n-1}}{f(\omega_n)} \frac{\delta^{n-1} \langle Q \rangle^{(n)}(\omega)}{\delta f(\omega_{n-1}) \cdots \delta f(\omega_1)},
\label{eq:sm_derivative_n-1}
\end{align}
as shown in Eq.~(4) in the main text.

\subsection{Equation (9)}
By directly calculating the first variation, we obtain Eq.~(9) as follows:
\begin{align}
\delta_{g} \langle Q \rangle^{(2)}(\omega)
&\equiv \langle Q \rangle^{(2)}_{f+\epsilon g}(\omega) - \langle Q \rangle^{(2)}_{f}(\omega) \notag \\
&= \int_{-\infty}^{\infty} \frac{\mathrm{d}\omega_1\, \mathrm{d}\omega_2}{2\pi}\, \delta(\omega_1+\omega_2-\omega)\, \chi^{(2)}(\omega_1,\omega_2) [f(\omega_1) + \epsilon g(\omega_1)] [f(\omega_2) + \epsilon g(\omega_2)] \notag \\
&\quad - \int_{-\infty}^{\infty} \frac{\mathrm{d}\omega_1\, \mathrm{d}\omega_2}{2\pi}\, \delta(\omega_1+\omega_2-\omega)\, \chi^{(2)}(\omega_1,\omega_2) f(\omega_1) f(\omega_2) \notag \\
&= \epsilon \int_{-\infty}^{\infty} \frac{\mathrm{d}\omega_1\, \mathrm{d}\omega_2}{2\pi}\, \delta(\omega_1+\omega_2-\omega)\, \chi^{(2)}(\omega_1,\omega_2) [f(\omega_1) g(\omega_2) + g(\omega_1) f(\omega_2)] + \mathcal{O}(\epsilon^2) \notag \\
&= \epsilon \int_{-\infty}^{\infty} \frac{\mathrm{d}\omega_1\, \mathrm{d}\omega_2}{2\pi}\, \delta(\omega_1+\omega_2-\omega) [\chi^{(2)}(\omega_2,\omega_1) f(\omega_2) g(\omega_1) + \chi^{(2)}(\omega_1,\omega_2) g(\omega_1) f(\omega_2)] + \mathcal{O}(\epsilon^2) \notag \\
&= \epsilon \int_{-\infty}^{\infty} \mathrm{d}\omega_1\, g(\omega_1) \int_{-\infty}^{\infty} \frac{\mathrm{d}\omega_2}{2\pi}\, \delta(\omega_1+\omega_2-\omega) [\chi^{(2)}(\omega_1, \omega_2) + \chi^{(2)}(\omega_2, \omega_1) ] f(\omega_2) + \mathcal{O}(\epsilon^2)\notag \\
&= \epsilon \int_{-\infty}^{\infty} \mathrm{d}\omega_1\, g(\omega_1) \frac{1}{2\pi} {\underbrace{[\chi^{(2)}(\omega_1, \omega-\omega_1) + \chi^{(2)}(\omega-\omega_1, \omega_1) ]}_{=\bar{\chi}^{(2)}(\omega_1, \omega-\omega_1)}} f(\omega-\omega_1) + \mathcal{O}(\epsilon^2) \notag \\
&= \epsilon \int_{-\infty}^{\infty} \mathrm{d}\omega_1\, g(\omega_1) \frac{\bar{\chi}^{(2)}(\omega_1,\omega-\omega_1) f(\omega-\omega_1)}{2\pi} + \mathcal{O}(\epsilon^2).
\end{align}
Note that $\chi^{(2)}$ appears in its symmetric form as $\bar{\chi}^{(2)}$ [see Eq.~(3)]; this also holds for higher order responses.

\section{Relationship to two-dimensional coherent spectroscopy}
In two-dimensional coherent spectroscopy (2DCS), two delta-function pulses are applied at time $0$ and $\tau\ (>0)$, and then a physical quantity $Q$ is measured at time $t+\tau\ (> t > 0)$.
For simplicity, here we omit the indices of the physical quantities and assume that $\langle Q \rangle^{(0)} = 0$.
The two delta-function pulses can be written as
\begin{align}
f(t) = F_0 \delta(t), \quad
g(t) = F_\tau \delta(t-\tau).
\end{align}
From Eq.~\eqref{eq:response_time}, we obtain the following expressions up to the third order:
\begin{align}
\langle Q \rangle_{f}(t+\tau) &\approx \int_{-\infty}^{\infty} \mathrm{d}t_1\, \chi^{(1)}(t+\tau;t_1) f(t_1)
+ \int_{-\infty}^{\infty} \mathrm{d}t_1 \mathrm{d}t_2\, \chi^{(2)}(t+\tau;t_1,t_2) f(t_1) f(t_2) \notag \\
&\quad + \int_{-\infty}^{\infty} \mathrm{d}t_1 \mathrm{d}t_2 \mathrm{d}t_3\, \chi^{(3)}(t+\tau;t_1,t_2,t_3) f(t_1) f(t_2) f(t_3) \notag \\
&= \chi^{(1)}(t+\tau;0) F_0
+ \chi^{(2)}(t+\tau;0,0) F_0^2
+ \chi^{(3)}(t+\tau;0,0,0) F_0^3, \\
\langle Q \rangle_{g}(t+\tau) &\approx \int_{-\infty}^{\infty} \mathrm{d}t_1\, \chi^{(1)}(t+\tau;t_1) g(t_1)
+ \int_{-\infty}^{\infty} \mathrm{d}t_1 \mathrm{d}t_2\, \chi^{(2)}(t+\tau;t_1,t_2) g(t_1) g(t_2) \notag \\
&\quad + \int_{-\infty}^{\infty} \mathrm{d}t_1 \mathrm{d}t_2 \mathrm{d}t_3\, \chi^{(3)}(t+\tau;t_1,t_2,t_3) g(t_1) g(t_2) g(t_3) \notag \\
&= \chi^{(1)}(t+\tau;\tau) F_\tau
+ \chi^{(2)}(t+\tau;\tau,\tau) F_\tau^2
+ \chi^{(3)}(t+\tau;\tau,\tau,\tau) F_\tau^3, \\
\langle Q \rangle_{f+g}(t+\tau) &\approx \int_{-\infty}^{\infty} \mathrm{d}t_1\, \chi^{(1)}(t+\tau;t_1) [f(t_1)+g(t_1)]
+ \int_{-\infty}^{\infty} \mathrm{d}t_1 \mathrm{d}t_2\, \chi^{(2)}(t+\tau;t_1,t_2) [f(t_1)+g(t_1)] [f(t_2)+g(t_2)] \notag \\
&\quad + \int_{-\infty}^{\infty} \mathrm{d}t_1 \mathrm{d}t_2 \mathrm{d}t_3\, \chi^{(3)}(t+\tau;t_1,t_2,t_3) [f(t_1)+g(t_1)] [f(t_2)+g(t_2)] [f(t_3)+g(t_3)] \notag \\
&= \chi^{(1)}(t+\tau;0) F_0 + \chi^{(1)}(t+\tau;\tau) F_\tau \notag \\
&\quad + \chi^{(2)}(t+\tau;0,0) F_0^2
+ \cancel{\chi^{(2)}(t+\tau;0,\tau)} F_0 F_\tau
+ \chi^{(2)}(t+\tau;\tau,0) F_\tau F_0
+ \chi^{(2)}(t+\tau;\tau,\tau) F_\tau^2 \notag \\
&\quad + \chi^{(3)}(t+\tau;0,0,0) F_0^3
+ \bigl[ \cancel{\chi^{(3)}(t+\tau;0,0,\tau)} + \cancel{\chi^{(3)}(t+\tau;0,\tau,0)} + \chi^{(3)}(t+\tau;\tau,0,0) \bigr] F_0^2 F_\tau \notag \\
&\quad + \bigl[ \cancel{\chi^{(3)}(t+\tau;0,\tau,\tau)} + \cancel{\chi^{(3)}(t+\tau;\tau,0,\tau)} + \chi^{(3)}(t+\tau;\tau,\tau,0) \bigr] F_0 F_\tau^2
+ \chi^{(3)}(t+\tau;\tau,\tau,\tau) F_\tau^3. \label{eq:2dcs_Qfg}
\end{align}
Therefore,
\begin{align}
\langle Q \rangle_{\text{NL}}(t+\tau)
&\equiv \langle Q \rangle_{f+g}(t+\tau)
- \langle Q \rangle_{f}(t+\tau)
- \langle Q \rangle_{g}(t+\tau) \notag \\
&\approx \chi^{(2)}(t+\tau;\tau,0) F_\tau F_0
+ \chi^{(3)}(t+\tau;\tau,0,0) F_0^2 F_\tau
+ \chi^{(3)}(t+\tau;\tau,\tau,0) F_0 F_\tau^2 \notag \\
&= \chi^{(2)}(t,t+\tau) F_\tau F_0
+ \chi^{(3)}(t,t+\tau,t+\tau) F_0^2 F_\tau
+ \chi^{(3)}(t,t,t+\tau) F_0 F_\tau^2.
\label{eq:2dcs_NL}
\end{align}
Here, we have used the causality condition, such as $\chi^{(2)}(t+\tau;0,\tau) = 0$ for $\tau>0$.
We can numerically extract the second- and third-order response functions in Eq.~\eqref{eq:2dcs_NL} by varying $F_0$ and $F_\tau$.
In the conventional 2DCS, the Fourier transformation with respect to $t$ and $\tau$ are performed.
For the second-order response, we find that
\begin{align}
\tilde{\chi}^{(2)}(\omega_t,\omega_\tau)
\equiv \int_{-\infty}^{\infty} \mathrm{d}t \mathrm{d}\tau\, \mathrm{e}^{\mathrm{i}\omega_t t+\mathrm{i}\omega_\tau \tau} \chi^{(2)}(t,t+\tau)
= \int_{-\infty}^{\infty} \mathrm{d}\bar{t}_1 \mathrm{d}\bar{t}_2\, \mathrm{e}^{\mathrm{i}(\omega_t-\omega_\tau)\bar{t}_1+\mathrm{i}\omega_\tau \bar{t}_2} \chi^{(2)}(\bar{t}_1,\bar{t}_2)
= \chi^{(2)}(\omega_t-\omega_\tau, \omega_\tau),
\end{align}
which indicates that 2DCS can provide $\chi^{(2)}(\omega_1, \omega_2)$ in Eq.~\eqref{eq:def_chi_nth} across the entire two-dimensional frequency space.
Note that when $\tau < 0$, $\chi^{(2)}(t+\tau; \tau,0) = \chi^{(2)}(t, t+\tau)$ vanishes, whereas the other second-order term, $\chi^{(2)}(t+\tau;0,\tau) = \chi^{(2)}(t+\tau,t)$, becomes nonzero.
In this case, the Fourier transformation yields:
\begin{align}
\tilde{\chi}^{(2')}(\omega_t, \omega_\tau)
\equiv \int_{-\infty}^{\infty} \mathrm{d}t \mathrm{d}\tau\, \mathrm{e}^{\mathrm{i}\omega_t t + \mathrm{i}\omega_\tau \tau} \chi^{(2)}(t+\tau,t)
= \int_{-\infty}^{\infty} \mathrm{d}\bar{t}_1 \mathrm{d}\bar{t}_2\, \mathrm{e}^{\mathrm{i}\omega_\tau \bar{t}_1 + \mathrm{i}(\omega_t - \omega_\tau)\bar{t}_2} \chi^{(2)}(\bar{t}_1, \bar{t}_2)
= \chi^{(2)}(\omega_\tau, \omega_t-\omega_\tau).
\end{align}
Thus, $\chi^{(2)}(t+\tau; \tau, 0)$ for $\tau > 0$ and $\chi^{(2)}(t+\tau; 0, \tau)$ for $\tau < 0$ contain equivalent information about $\chi^{(2)}(\omega_1, \omega_2)$.

In contrast, for the third-order response, it is shown that
\begin{align}
\tilde{\chi}^{(3,1)}(\omega_t,\omega_\tau)
&\equiv \int_{-\infty}^{\infty} \mathrm{d}t \mathrm{d}\tau\, \mathrm{e}^{\mathrm{i}\omega_t t+\mathrm{i}\omega_\tau \tau} \chi^{(3)}(t,t+\tau,t+\tau) \notag \\
&= \int_{-\infty}^{\infty} \mathrm{d}t \mathrm{d}\tau \mathrm{d}\tau'\, \mathrm{e}^{\mathrm{i}\omega_t t+\mathrm{i}\omega_\tau \tau} \delta(\tau'-\tau) \chi^{(3)}(t,t+\tau,t+\tau') \notag \\
&= \int_{-\infty}^{\infty} \frac{\mathrm{d}\omega'}{2\pi} \int_{-\infty}^{\infty} \mathrm{d}t \mathrm{d}\tau \mathrm{d}\tau'\, \mathrm{e}^{\mathrm{i}\omega_t t+\mathrm{i}\omega_\tau \tau+\mathrm{i}\omega'(\tau'-\tau)} \chi^{(3)}(t,t+\tau,t+\tau') \notag \\
&= \int_{-\infty}^{\infty} \frac{\mathrm{d}\omega'}{2\pi} \int_{-\infty}^{\infty} \mathrm{d}\bar{t}_1 \mathrm{d}\bar{t}_2 \mathrm{d}\bar{t}_3\, \mathrm{e}^{\mathrm{i}(\omega_t-\omega_\tau)\bar{t}_1 + \mathrm{i}(\omega_\tau-\omega')\bar{t}_2 + \mathrm{i} \omega' \bar{t}_3} \chi^{(3)}(\bar{t}_1,\bar{t}_2,\bar{t}_3) \notag \\
&= \int_{-\infty}^{\infty} \frac{\mathrm{d}\omega'}{2\pi}\, \chi^{(3)}(\omega_t-\omega_\tau, \omega_\tau-\omega',\omega'), \label{eq:2dcs_chi31}
\end{align}
and
\begin{align}
\tilde{\chi}^{(3,2)}(\omega_t,\omega_\tau)
&\equiv \int_{-\infty}^{\infty} \mathrm{d}t \mathrm{d}\tau\, \mathrm{e}^{\mathrm{i}\omega_t t+\mathrm{i}\omega_\tau \tau} \chi^{(3)}(t,t,t+\tau) \notag \\
&= \int_{-\infty}^{\infty} \mathrm{d}t \mathrm{d}t' \mathrm{d}\tau\, \mathrm{e}^{\mathrm{i}\omega_t t+\mathrm{i}\omega_\tau \tau} \delta(t'-t) \chi^{(3)}(t,t',t'+\tau) \notag \\
&= \int_{-\infty}^{\infty} \frac{\mathrm{d}\omega'}{2\pi} \int_{-\infty}^{\infty} \mathrm{d}t \mathrm{d}t' \mathrm{d}\tau\, \mathrm{e}^{\mathrm{i}\omega_t t + \mathrm{i}\omega_\tau \tau + \mathrm{i} \omega'(t'-t)} \chi^{(3)}(t,t',t'+\tau) \notag \\
&= \int_{-\infty}^{\infty} \frac{\mathrm{d}\omega'}{2\pi} \int_{-\infty}^{\infty} \mathrm{d}\bar{t}_1 \mathrm{d}\bar{t}_2 \mathrm{d}\bar{t}_3\, \mathrm{e}^{\mathrm{i}(\omega_t-\omega')\bar{t}_1 + \mathrm{i}(\omega'-\omega_\tau)\bar{t}_2 + \mathrm{i}\omega_\tau \bar{t}_3} \chi^{(3)}(\bar{t}_1, \bar{t}_2, \bar{t}_3) \notag \\
&= \int_{-\infty}^{\infty} \frac{\mathrm{d}\omega'}{2\pi}\, \chi^{(3)}(\omega_t-\omega', \omega'-\omega_\tau, \omega_\tau). \label{eq:2dcs_chi32}
\end{align}
Equations~\eqref{eq:2dcs_chi31} and \eqref{eq:2dcs_chi32} indicate that, for the third and higher orders, only the \textit{partial} response functions can be obtained through 2DCS.
Although the \textit{full} $n$th-order response functions can, in principle, be retrieved by $n$-dimensional spectroscopy, the computational cost increases significantly because of the need to sweep the time delays between $n$ delta-function pulses.

\section{Response functions in quantum systems} \label{sec:response_in_quantum_systems}
Suppose that the Hamiltonian of a quantum system at time $t$ is given by
\begin{align}
\mathcal{H}(t) = \mathcal{H}_0 + \mathcal{V}(t), \quad
\mathcal{V}(t) = -\sum_{\beta} f_{\beta}(t) Q_{\beta},
\label{eq:response_hamiltonian}
\end{align}
where $Q_{\beta}$ is a physical quantity labeled by $\beta$, and $f_{\beta}(t)$ is an external field linearly coupled to $Q_{\beta}$.
The interaction-picture operator of $Q_{\alpha}$ is defined by
\begin{align}
\hat{Q}_{\alpha}(t) = U_0(t_0,t) Q_{\alpha} U_0(t,t_0), \quad
U_0(t,t_0) = \mathrm{e}^{-\mathrm{i}\mathcal{H}_0\cdot (t-t_0)}.
\end{align}
The time evolution of the many-body density matrix $\rho(t)$ is governed by the von Neumann equation,
\begin{align}
\mathrm{i} \partial_t \rho(t) = [\mathcal{H}(t), \rho(t)],
\end{align}
and its solution can be formally written as
\begin{align}
\rho(t) = U(t,t_0) \rho(t_0) U(t_0,t), \quad
U(t,t_0) = \mathcal{T}_{\mathcal{C}(t,t_0)} \exp\left[-\mathrm{i}\int_{t_0}^{t} \mathrm{d}\bar{t}\, \mathcal{H}(\bar{t})\right],
\end{align}
where $\mathcal{T}_{\mathcal{C}(t,t_0)}$ represents time-ordering along the contour from time $t_0$ to $t$.
Here, we introduce the $S$-matrix:
\begin{align}
S(t,t_0) = U_0(t,t_0)^\dagger U(t,t_0)
= \mathcal{T}_{\mathcal{C}(t,t_0)} \exp\left[-\mathrm{i}\int_{t_0}^{t} \mathrm{d}\bar{t}\, \hat{\mathcal{V}}(\bar{t}) \right],
\end{align}
and define the density matrix in the interaction picture as
\begin{align}
\hat{\rho}(t) = S(t,t_0) \rho(t_0) S(t_0,t),
\end{align}
which satisfies the following equation:
\begin{align}
\mathrm{i} \partial_t \hat{\rho}(t)
= [\hat{\mathcal{V}}(t), \hat{\rho}(t)].
\label{eq:vonNeumann_int}
\end{align}
The successive integration of Eq.~\eqref{eq:vonNeumann_int} yields
\begin{align}
\hat{\rho}(t) = \rho_0 + \sum_{n=1}^{\infty} (-\mathrm{i})^n \int_{t_0}^{t} \mathrm{d}t_1 \int_{t_0}^{t_1} \mathrm{d}t_2 \cdots \int_{t_0}^{t_{n-1}} \mathrm{d}t_{n}\, [\hat{\mathcal{V}}(t_1), [\hat{\mathcal{V}}(t_2), \cdots [\hat{\mathcal{V}}(t_n), \rho_0] \cdots ]],
\end{align}
where $\rho_0 = \hat{\rho}(t_0) = \rho(t_0)$.
Then, we obtain
\begin{gather}
\langle Q_{\alpha} \rangle(t)
= \Tr[\hat{\rho}(t) \hat{Q}_{\alpha}(t)]
= \langle Q_{\alpha}\rangle_0 + \sum_{n=1}^{\infty} \langle Q_{\alpha} \rangle^{(n)}(t), \\
\langle Q_{\alpha} \rangle^{(n)}(t)
= (-\mathrm{i})^n \int_{t_0}^{t} \mathrm{d}t_1 \int_{t_0}^{t_1} \mathrm{d}t_2 \cdots \int_{t_0}^{t_{n-1}} \mathrm{d}t_n\, \left\langle [\cdots [[\hat{Q}_{\alpha}(t), \hat{\mathcal{V}}(t_1)], \hat{\mathcal{V}}(t_2)], \cdots \hat{\mathcal{V}}(t_n)] \right\rangle_0,
\label{eq:Q_nth}
\end{gather}
with $\langle \bullet \rangle_0 = \langle \bullet \rangle^{(0)} = \Tr[\rho_0\, \bullet]$, where we have iteratively used the formula: $\Tr([A,B]C) = \Tr(B[C,A])$.
By taking the initial time to be $t_0 \to -\infty$ and substituting Eq.~\eqref{eq:response_hamiltonian}, Eq.~\eqref{eq:Q_nth} can be written as
\begin{align}
\langle Q_{\alpha} \rangle^{(n)}(t)
= \sum_{\{\beta_i\}} \int_{-\infty}^{\infty} \mathrm{d}t_1 \cdots \mathrm{d}t_n\, \chi_{\alpha\beta_1\cdots\beta_n}^{(n)}(t;t_1,\dots,t_n) f_{\beta_1}(t_1) \cdots f_{\beta_n}(t_n),
\label{eq:response_nth_time}
\end{align}
where $\chi_{\alpha\beta_1\cdots\beta_n}^{(n)}(t;t_1,\dots,t_n)$ is the $n$th-order response function defined by
\begin{align}
\chi_{\alpha\beta_1\cdots\beta_n}^{(n)}(t;t_1,\dots,t_n)
= \mathrm{i}^n \varTheta(t-t_1) \varTheta(t_1-t_2) \cdots \varTheta(t_{n-1}-t_{n}) \left\langle [\cdots [[\hat{Q}_{\alpha}(t), \hat{Q}_{\beta_1}(t_1)], \hat{Q}_{\beta_2}(t_2)], \cdots \hat{Q}_{\beta_n}(t_n)] \right\rangle_0.
\label{eq:response_nth_quantum}
\end{align}
Here, $\varTheta$ denotes the unit step function, ensuring causality.

Let us consider a free fermion system whose Hamiltonian $\mathcal{H}_0$ is given by
\begin{align}
\mathcal{H}_0 = \sum_{ij} h_{ij} c_i^\dagger c_{j} = \sum_{\nu} \varepsilon_\nu a_{\nu}^\dagger a_{\nu},
\end{align}
where $c_i^\dagger$ and $a_\nu^\dagger$ are the creation operators for fermions, and $\varepsilon_\nu$ denotes the one-body energy.
In the following, we derive the expressions for the first-, second-, and third-order response of a physical quantity,
\begin{align}
Q_{\alpha} = \sum_{ij} Q_{\alpha}^{ij} c_{i}^\dagger c_j = \sum_{\mu\nu} Q_{\alpha}^{\mu\nu} a_{\mu}^\dagger a_{\nu},
\end{align}
assuming that $Q_{\alpha}$ is time-independent in the Schr\"odinger picture.
Using the commutation relations:
\begin{gather}
[a_\mu^\dagger a_\nu, a_{\mu'}^\dagger a_{\nu'}] = \delta_{\nu\mu'} a_{\mu}^\dagger a_{\nu'} - \delta_{\mu\nu'} a_{\mu'}^\dagger a_{\nu}, \\
[a_\mu^\dagger a_\nu, \mathcal{H}_0] = (\varepsilon_\nu-\varepsilon_\mu) a_{\mu}^\dagger a_{\nu},
\end{gather}
we obtain the expression for the first-order response,
\begin{gather}
\chi_{\alpha\beta}^{(1)}(t;t_1)
= \mathrm{i} \varTheta(t-t_1) \sum_{\mu\nu} Q_{\alpha}^{\mu\nu} Q_{\beta}^{\nu\mu} \mathrm{e}^{-\mathrm{i}(\varepsilon_{\nu}-\varepsilon_{\mu})(t-t_1)} (n_\mu-n_\nu) = \chi_{\alpha\beta}^{(1)}(\bar{t}_1), \\
\chi_{\alpha\beta}^{(1)}(\omega)
= \int_{-\infty}^{\infty} \mathrm{d}\bar{t}_1\, \mathrm{e}^{\mathrm{i}\omega^+ \bar{t}_1} \chi_{\alpha\beta}^{(1)}(\bar{t}_1)
= \sum_{\mu\nu} Q_{\alpha}^{\mu\nu} Q_{\beta}^{\nu\mu} \frac{n_\nu-n_\mu}{\omega^+ -(\varepsilon_\nu-\varepsilon_\mu)},
\label{eq:chi_1st}
\end{gather}
where $n_\nu = \langle a_{\nu}^\dagger a_{\nu} \rangle_0$ corresponds to the Fermi--Dirac distribution function, and $\omega^+$ denotes $\omega$ with a small positive imaginary part $\eta$, i.e., $\omega^+ = \omega + \mathrm{i} \eta$.
Similarly, the second-order response is derived as follows:
\begin{align}
\chi_{\alpha\beta_1\beta_2}^{(2)}(\omega_1,\omega_2)
= \sum_{\mu_i\nu_i} \Biggl[ \frac{Q_{\alpha}^{\mu_0\nu_0} Q_{\beta_1}^{\nu_0\nu_1} Q_{\beta_2}^{\nu_1\mu_0} (n_{\mu_0}-n_{\nu_1})}{(\omega_1^+ + \omega_2^+ -\varepsilon_{\nu_0} + \varepsilon_{\mu_0}) (\omega_2^+ -\varepsilon_{\nu_1} +\varepsilon_{\mu_0})} + \frac{Q_{\alpha}^{\mu_0\nu_0} Q_{\beta_2}^{\nu_0\mu_1} Q_{\beta_1}^{\mu_1\mu_0} (n_{\nu_0} - n_{\mu_1})}{(\omega_1^+ + \omega_2^+ -\varepsilon_{\nu_0} +\varepsilon_{\mu_0}) (\omega_2^+ -\varepsilon_{\nu_0} + \varepsilon_{\mu_1})} \Biggr],
\label{eq:chi_2nd}
\end{align}
and the third-order response can be expressed as:
\begin{align}
\chi_{\alpha\beta_1\beta_2\beta_3}^{(3)}(\omega_1,\omega_2,\omega_3)
&= \sum_{\mu_i\nu_i} \Biggl[
\frac{Q_{\alpha}^{\mu_0\nu_0} Q_{\beta_1}^{\nu_0\nu_1} Q_{\beta_2}^{\nu_1\nu_2} Q_{\beta_3}^{\nu_2\mu_0} (n_{\nu_2} - n_{\mu_0})}{(\omega_1^+ + \omega_2^+ + \omega_3^+ -\varepsilon_{\nu_0} +\varepsilon_{\mu_0}) (\omega_2^+ + \omega_3^+ -\varepsilon_{\nu_1} +\varepsilon_{\mu_0}) (\omega_3^+ -\varepsilon_{\nu_2} +\varepsilon_{\mu_0})} \notag \\
&\quad + \frac{Q_{\alpha}^{\mu_0\nu_0} Q_{\beta_1}^{\nu_0\nu_1} Q_{\beta_3}^{\nu_1\mu_2} Q_{\beta_2}^{\mu_2\mu_0} (n_{\mu_2} - n_{\nu_1})}{(\omega_1^+ + \omega_2^+ + \omega_3^+ -\varepsilon_{\nu_0} +\varepsilon_{\mu_0}) (\omega_2^+ + \omega_3^+ -\varepsilon_{\nu_1} +\varepsilon_{\mu_0}) (\omega_3^+ -\varepsilon_{\nu_1} +\varepsilon_{\mu_2})} \notag \\
&\quad + \frac{Q_{\alpha}^{\mu_0\nu_0} Q_{\beta_2}^{\nu_0\nu_2} Q_{\beta_3}^{\nu_2\mu_1} Q_{\beta_1}^{\mu_1\mu_0} (n_{\mu_1} - n_{\nu_2})}{(\omega_1^+ + \omega_2^+ + \omega_3^+ -\varepsilon_{\nu_0} +\varepsilon_{\mu_0}) (\omega_2^+ + \omega_3^+ -\varepsilon_{\nu_0} +\varepsilon_{\mu_1}) (\omega_3^+ -\varepsilon_{\nu_2} +\varepsilon_{\mu_1})} \notag \\
&\quad + \frac{Q_{\alpha}^{\mu_0\nu_0} Q_{\beta_3}^{\nu_0\mu_2} Q_{\beta_2}^{\mu_2\mu_1} Q_{\beta_1}^{\mu_1\mu_0} (n_{\nu_0} - n_{\mu_2})}{(\omega_1^+ + \omega_2^+ + \omega_3^+ -\varepsilon_{\nu_0} +\varepsilon_{\mu_0}) (\omega_2^+ + \omega_3^+ -\varepsilon_{\nu_0} +\varepsilon_{\mu_1}) (\omega_3^+ -\varepsilon_{\nu_0} +\varepsilon_{\mu_2})}
\Biggr].
\label{eq:chi_3rd}
\end{align}
Note that these higher-order response functions need to be symmetrized as follows:
\begin{gather}
\bar{\chi}_{\alpha\beta_1\beta_2}^{(2)}(\omega_1,\omega_2)
= \sum_{\sigma\in\mathfrak{S}_2} \chi_{\alpha\beta_{\sigma(1)}\beta_{\sigma(2)}}^{(2)}(\omega_{\sigma(1)},\omega_{\sigma(2)})
= \chi_{\alpha\beta_1\beta_2}^{(2)}(\omega_1,\omega_2)
+ \chi_{\alpha\beta_2\beta_1}^{(2)}(\omega_2,\omega_1), \label{eq:chi2_sym} \\
\bar{\chi}_{\alpha\beta_1\beta_2\beta_3}^{(3)}(\omega_1,\omega_2,\omega_3)
= \sum_{\sigma\in\mathfrak{S}_3} \chi_{\alpha\beta_{\sigma(1)}\beta_{\sigma(2)}\beta_{\sigma(3)}}^{(3)}(\omega_{\sigma(1)},\omega_{\sigma(2)},\omega_{\sigma(3)})
= (\text{6 terms from all permutations}). \label{eq:chi3_sym}
\end{gather}

\section{Nonlinear optical responses in the velocity gauge}
When introducing the vector potential $\bm{A}$ through Peierls substitution, the Hamiltonian of a free fermion system takes the following form:
\begin{align}
\mathcal{H}[\bm{A}] = \sum_{\bm{k}} \sum_{ij} h_{ij}(\bm{k}-\bm{A}) c_{\bm{k}i}^\dagger c_{\bm{k}j},
\end{align}
where $\bm{k}$ denotes the crystal momentum, and $i$ and $j$ are indices representing internal degrees of freedom, such as spin, orbital, and sublattice.
This Hamiltonian describes a nonlinear coupling between the vector potential $\bm{A}$ and the electric current operator $\bm{J} = -\partial \mathcal{H}[\bm{A}]/\partial \bm{A}$.
In this case, the results from the previous section cannot be directly applied because the current operator $\bm{J}$ itself depends on the external field $\bm{A}$.
Therefore, $\bm{J}$ should be expanded in terms of $\bm{A}$ as follows:
\begin{align}
J_{\alpha} = -\frac{\partial \mathcal{H}[\bm{A}]}{\partial A_\alpha} = \sum_{n=0}^{\infty} \sum_{\{\beta_i\}} J_{\alpha\beta_1 \cdots \beta_n}^{(n)} A_{\beta_1} \cdots A_{\beta_{n}}, \quad
J_{\beta_0 \beta_1 \cdots \beta_n}^{(n)} = \frac{(-1)^n}{n!} \sum_{\bm{k}ij} \partial_{\beta_0} \partial_{\beta_1} \cdots \partial_{\beta_n} h_{ij}(\bm{k}) c_{\bm{k}i}^\dagger c_{\bm{k}j},
\end{align}
where $\partial_{\beta} = \partial/\partial k_{\beta}$.
Note that $J_{\beta_0 \beta_1 \cdots \beta_n}^{(n)}$ is invariant under any permutation of its indices.
By using this expansion, the Hamiltonian can be expressed as
\begin{align}
\mathcal{H}[\bm{A}] = \mathcal{H}[0] - \sum_{n=0}^{\infty} \sum_{\{\beta_i\}} \frac{A_{\beta_0} \cdots A_{\beta_n}}{n+1} J_{\beta_0 \cdots \beta_n}^{(n)}.
\end{align}
Regarding the second term as a coupling term with $\bm{A}(t)$, we consider the response of the electric current density at time $t$:
\begin{align}
\langle J_{\alpha} \rangle(t) = \frac{1}{V} \sum_{n=0}^{\infty} \sum_{\{\beta_i\}} \langle J_{\alpha\beta_1 \cdots \beta_n}^{(n)} \rangle(t) A_{\beta_1}(t) \cdots A_{\beta_n}(t),
\end{align}
where $V$ represents the volume of the system.
Since many types of physical quantities $\{J_{\alpha\beta_1\cdots\beta_n}^{(n)}\}$ and external fields $\{A_{\beta_0} \cdots A_{\beta_n}\}$ appear, we introduce the following notation for the $n$th-order response function:
\begin{align}
&\langle\!\langle Q_{\alpha}(t); Q_{\beta_1}(t_1), \dots, Q_{\beta_n}(t_n) \rangle\!\rangle
= \langle\!\langle Q_{\alpha}; Q_{\beta_1}, \dots, Q_{\beta_n} \rangle\!\rangle(\bar{t}_1,\dots,\bar{t}_n) \notag \\
&= \mathrm{i}^n \varTheta(t-t_1) \varTheta(t_1-t_2) \cdots \varTheta(t_{n-1}-t_n) \left\langle [\cdots [[\hat{Q}_{\alpha}(t), \hat{Q}_{\beta_1}(t_1)], \hat{Q}_{\beta_2}(t_2)], \cdots \hat{Q}_{\beta_n}(t_n)] \right\rangle_0
\end{align}
in the time domain, and
\begin{align}
\langle\!\langle Q_{\alpha}; Q_{\beta_1}, \dots, Q_{\beta_n} \rangle\!\rangle(\omega_1,\dots,\omega_n)
= \int_{-\infty}^{\infty} \mathrm{d}\bar{t}_1 \cdots \mathrm{d}\bar{t}_n\, \mathrm{e}^{\mathrm{i}\omega_1 \bar{t}_1 + \cdots + \mathrm{i}\omega_n \bar{t}_n} \langle\!\langle Q_{\alpha}; Q_{\beta_1}, \dots, Q_{\beta_n} \rangle\!\rangle(\bar{t}_1,\dots,\bar{t}_n)
\end{align}
in the frequency domain.

The first-order response is given by
\begin{align}
\langle J_{\alpha} \rangle^{(1)}(t)
= \frac{1}{V} \sum_{\beta_1} \left[ \langle J_{\alpha\beta_1}^{(1)} \rangle_0 A_{\beta_1}(t)
+ \int_{-\infty}^{\infty} \mathrm{d}t_1\, \langle\!\langle J_{\alpha}^{(0)}(t); J_{\beta_1}^{(0)}(t_1) \rangle\!\rangle A_{\beta_1}(t_1) \right].
\label{eq:current_1st_time}
\end{align}
Here, the first term represents the ``zeroth-order'' response of the diamagnetic current $J_{\alpha\beta_1}^{(1)} A_{\beta_1}$, and the second term represents the first-order response of the paramagnetic current $J_{\alpha}^{(0)}$ to the vector potential $\bm{A}$.
The Fourier transformation of Eq.~\eqref{eq:current_1st_time} yields
\begin{align}
\langle J_{\alpha} \rangle^{(1)}(\omega)
= \sum_{\beta_1} \chi_{\alpha\beta_1}^{(1)}(\omega) A_{\beta_1}(\omega), \quad
V \chi_{\alpha\beta_1}^{(1)}(\omega)
= \langle J_{\alpha\beta_1}^{(1)} \rangle_0 + \langle\!\langle J_{\alpha}^{(0)}; J_{\beta_1}^{(0)} \rangle\!\rangle(\omega).
\end{align}

The second-order response consists of the following four terms:
\begin{align}
\langle J_{\alpha} \rangle^{(2)}(t)
&= \frac{1}{V} \sum_{\beta_1\beta_2} \biggl[
\langle J_{\alpha\beta_1\beta_2}^{(2)} \rangle_0 A_{\beta_1}(t) A_{\beta_2}(t)
+ \int_{-\infty}^{\infty} \mathrm{d}t_1 \mathrm{d}t_2\, \langle\!\langle J_{\alpha}^{(0)}(t); J_{\beta_1}^{(0)}(t_1), J_{\beta_2}^{(0)}(t_2) \rangle\!\rangle A_{\beta_1}(t_1) A_{\beta_2}(t_2) \notag \\
&\quad + \int_{-\infty}^{\infty} \mathrm{d}t_1\, \langle\!\langle J_{\alpha}^{(0)}(t); J_{\beta_1\beta_2}^{(1)}(t_1) \rangle\!\rangle \frac{A_{\beta_1}(t_1) A_{\beta_2}(t_2)}{2}
+ \int_{-\infty}^{\infty} \mathrm{d}t_1\, \langle\!\langle J_{\alpha\beta_1}^{(1)}(t); J_{\beta_2}^{(0)}(t_1) \rangle\!\rangle A_{\beta_1}(t) A_{\beta_2}(t_1)
\biggr].
\end{align}
Here, the first term is the zeroth-order response of $J_{\alpha\beta_1\beta_2}^{(2)} A_{\beta_1} A_{\beta_2}$; the second term represents the second-order response of $J_{\alpha}^{(0)}$ to $\bm{A}$; the third term denotes $J_{\alpha}^{(0)}$ induced by $J_{\beta_1\beta_2}^{(1)} A_{\beta_1} A_{\beta_2}/2$; and the fourth term corresponds to $J_{\alpha\beta_1}^{(1)} A_{\beta_1}$ induced by $J_{\beta_2}^{(0)} A_{\beta_2}$.
By performing the Fourier transformation, we obtain
\begin{align}
\langle J_{\alpha} \rangle^{(2)}(\omega)
= \sum_{\beta_1\beta_2} \int_{-\infty}^{\infty} \frac{\mathrm{d}\omega_1 \mathrm{d}\omega_2}{2\pi}\, \delta(\omega_1+\omega_2-\omega) \chi_{\alpha\beta_1\beta_2}^{(2)}(\omega_1,\omega_2) A_{\beta_1}(\omega_1) A_{\beta_2}(\omega_2),
\end{align}
where
\begin{align}
V \chi_{\alpha\beta_1\beta_2}^{(2)}(\omega_1,\omega_2)
= \langle J_{\alpha\beta_1\beta_2}^{(2)} \rangle_0
+ \langle\!\langle J_{\alpha}^{(0)}; J_{\beta_1}^{(0)}, J_{\beta_2}^{(0)} \rangle\!\rangle(\omega_1,\omega_2)
+ \frac{1}{2} \langle\!\langle J_{\alpha}^{(0)}; J_{\beta_1\beta_2}^{(1)} \rangle\!\rangle(\omega_1+\omega_2)
+ \langle\!\langle J_{\alpha\beta_1}^{(1)}; J_{\beta_2}^{(0)} \rangle\!\rangle(\omega_2). \label{eq:chi2_optical}
\end{align}
Similarly, the third-order response is expressed as
\begin{align}
\langle J_{\alpha} \rangle^{(3)}(\omega)
= \sum_{\beta_1\beta_2\beta_3} \int_{-\infty}^{\infty} \frac{\mathrm{d}\omega_1 \mathrm{d}\omega_2 \mathrm{d}\omega_3}{(2\pi)^2}\, \delta(\omega_1+\omega_2+\omega_3-\omega) \chi_{\alpha\beta_1\beta_2\beta_3}^{(3)}(\omega_1,\omega_2,\omega_3) A_{\beta_1}(\omega_1) A_{\beta_2}(\omega_2) A_{\beta_3}(\omega_3),
\end{align}
where
\begin{align}
&V \chi_{\alpha\beta_1\beta_2\beta_3}^{(3)}(\omega_1,\omega_2,\omega_3) \notag \\
&= \langle J_{\alpha\beta_1\beta_2\beta_3}^{(3)} \rangle_0
+ \langle\!\langle J_{\alpha}^{(0)}; J_{\beta_1}^{(0)}, J_{\beta_2}^{(0)}, J_{\beta_3}^{(0)} \rangle\!\rangle(\omega_1,\omega_2,\omega_3) \notag \\
&\quad + \frac{1}{2} \langle\!\langle J_{\alpha}^{(0)}; J_{\beta_1}^{(0)}, J_{\beta_2\beta_3}^{(1)} \rangle\!\rangle(\omega_1, \omega_2+\omega_3)
+ \frac{1}{2} \langle\!\langle J_{\alpha}^{(0)}; J_{\beta_1\beta_2}^{(1)}, J_{\beta_3}^{(0)} \rangle\!\rangle(\omega_1+\omega_2, \omega_3)
+ \langle\!\langle J_{\alpha\beta_1}^{(1)}; J_{\beta_2}^{(0)}, J_{\beta_3}^{(0)} \rangle\!\rangle(\omega_2,\omega_3) \notag \\
&\quad + \frac{1}{3} \langle\!\langle J_{\alpha}^{(0)}; J_{\beta_1\beta_2\beta_3}^{(2)} \rangle\!\rangle(\omega_1+\omega_2+\omega_3)
+ \frac{1}{2} \langle\!\langle J_{\alpha\beta_1}^{(1)}; J_{\beta_2\beta_3}^{(1)} \rangle\!\rangle(\omega_2+\omega_3)
+ \langle\!\langle J_{\alpha\beta_1\beta_2}^{(2)}; J_{\beta_3}^{(0)} \rangle\!\rangle(\omega_3). \label{eq:chi3_optical}
\end{align}
We can calculate each term in Eqs.~\eqref{eq:chi2_optical} and \eqref{eq:chi3_optical} by replacing $\{Q_{\beta}\}$ in Eqs.~\eqref{eq:chi_1st}--\eqref{eq:chi_3rd} with the corresponding current operators $\{J_{\beta_0\cdots\beta_n}^{(n)}\}$.
Additionally, these second- and third-order response functions must be symmetrized in accordance with Eqs.~\eqref{eq:chi2_sym} and \eqref{eq:chi3_sym}.

\section{Hartree--Fock analysis of the Rice--Mele--Hubbard model}
In the Hartree--Fock (HF) approximation, the electron--electron interaction term in the Rice--Mele--Hubbard model [Eq.~(A1) in the main text] is decoupled as
\begin{align}
c_{j\uparrow}^\dagger c_{j\uparrow} c_{j\downarrow}^\dagger c_{j\downarrow}
&\approx c_{j\uparrow}^\dagger c_{j\uparrow} \langle c_{j\downarrow}^\dagger c_{j\downarrow} \rangle_0
+ \langle c_{j\uparrow}^\dagger c_{j\uparrow} \rangle_0 c_{j\downarrow}^\dagger c_{j\downarrow}
- \langle c_{j\uparrow}^\dagger c_{j\uparrow} \rangle_0 \langle c_{j\downarrow}^\dagger c_{j\downarrow} \rangle_0 \notag \\
&\quad - c_{j\uparrow}^\dagger  c_{j\downarrow} \langle c_{j\downarrow}^\dagger c_{j\uparrow} \rangle_0
- \langle c_{j\uparrow}^\dagger  c_{j\downarrow} \rangle_0 c_{j\downarrow}^\dagger c_{j\uparrow}
+ \langle c_{j\uparrow}^\dagger  c_{j\downarrow} \rangle_0 \langle c_{j\downarrow}^\dagger c_{j\uparrow} \rangle_0.
\end{align}
Here, the Fock terms, $\langle c_{j\downarrow}^\dagger c_{j\uparrow} \rangle_0$ and $\langle c_{j\uparrow}^\dagger  c_{j\downarrow} \rangle_0$, serve only to restore spin rotational symmetry, and thus these can be omitted without loss of generality.
The resulting one-body Hamiltonian takes the form:
\begin{align}
\mathcal{H}_{\mathrm{RMH}}^{\mathrm{HF}}
&= \mathcal{H}_{\mathrm{RMH}}^{\mathrm{HF},\uparrow} + \mathcal{H}_{\mathrm{RMH}}^{\mathrm{HF},\downarrow}
+ U \sum_{j} \left( \frac{1}{4} - \langle n_{j\uparrow} \rangle_0 \langle n_{j\downarrow} \rangle_0 \right), \\
\mathcal{H}_{\mathrm{RMH}}^{\mathrm{HF},\uparrow}
&= - \sum_{j} \frac{h_x - (-1)^{j} h_y}{2} \left( c_{j+1,\uparrow}^\dagger c_{j\uparrow} + \mathrm{H.c.} \right) - h_z \sum_{j} (-1)^j n_{j\uparrow}
+ U \sum_{j} \left( \langle n_{j\downarrow} \rangle_0 - \frac{1}{2} \right) n_{j\uparrow}, \\
\mathcal{H}_{\mathrm{RMH}}^{\mathrm{HF},\downarrow}
&= - \sum_{j} \frac{h_x - (-1)^{j} h_y}{2} \left( c_{j+1,\downarrow}^\dagger c_{j\downarrow} + \mathrm{H.c.} \right) - h_z \sum_{j} (-1)^j n_{j\downarrow}
+ U \sum_{j} \left( \langle n_{j\uparrow} \rangle_0 - \frac{1}{2} \right) n_{j\downarrow},
\end{align}
where $n_{js} = c_{js}^\dagger c_{js}$ denotes the number operator for an electron with real spin $s$ at site $j$.
Assuming the two-sublattice structure, we obtain the Fourier transform of $\mathcal{H}_{\mathrm{RMH}}^{\mathrm{HF}}$ as
\begin{align}
\mathcal{H}_{\mathrm{RMH}}^{\mathrm{HF}}
&= \sum_{k} \tilde{\varPsi}_k^\dagger \tilde{H}_{k} \tilde{\varPsi}_k + U \sum_{j} \left( \frac{1}{4} - \langle n_{j\uparrow} \rangle_0 \langle n_{j\downarrow} \rangle_0 \right), \\
\tilde{H}_{k}
&= -\left( \frac{U}{2} - \frac{U}{2} \langle \tau_0^\downarrow \rangle_0 \right) \tau_0^\uparrow
- h_x \cos(k/2) \tau_x^\uparrow - h_y \sin(k/2) \tau_y^\uparrow
- \left( h_z - \frac{U}{2} \langle \tau_z^\downarrow \rangle_0 \right) \tau_z^\uparrow \notag \\
&\quad -\left( \frac{U}{2} - \frac{U}{2} \langle \tau_0^\uparrow \rangle_0 \right) \tau_0^\downarrow
- h_x \cos(k/2) \tau_x^\downarrow - h_y \sin(k/2) \tau_y^\downarrow
- \left( h_z - \frac{U}{2} \langle \tau_z^\uparrow \rangle_0 \right) \tau_z^\downarrow.
\label{eq:rmh_hf_k}
\end{align}
Here, $\tau_\alpha^{\uparrow/\downarrow}$ is a $4\times 4$ matrix defined by
$
\tau_\alpha^{\uparrow/\downarrow} = (\sigma_0 \pm \sigma_z)/2 \otimes \tau_{\alpha} \ (\alpha=0,x,y,z),
$
and $\sigma_{\alpha}$ and $\tau_{\alpha}$ represent the Pauli matrices (with the identity matrices for $\alpha=0$) associated with the real spin and sublattice pseudospin, respectively.
Equation~\eqref{eq:rmh_hf_k} indicates that, in the HF approximation, the Hamiltonian reduces to two decoupled Rice--Mele models---one for each spin component---and that only the parameter $h_z$ is renormalized as $h_z \to \tilde{h}_z^{s}$, where
\begin{align}
\tilde{h}_z^{\uparrow} = h_z - \frac{U}{2} \langle \tau_z^{\downarrow} \rangle_0, \quad
\tilde{h}_z^{\downarrow} = h_z - \frac{U}{2} \langle \tau_z^{\uparrow} \rangle_0.
\label{eq:hz_renorm}
\end{align}
Therefore, when the deviation of $\tilde{h}_z^{s}$ from $h_z$ is small, the second-order response function $\chi^{(2)}(\omega_1,\omega_2)$ remains almost unchanged.
In contrast, a sign reversal in $\tilde{h}_z^{s}$ leads to a global sign flip in $\chi^{(2)}(\omega_1,\omega_2)$ across the entire frequency space.

\begin{figure}[b]\centering
\includegraphics[scale=1.33]{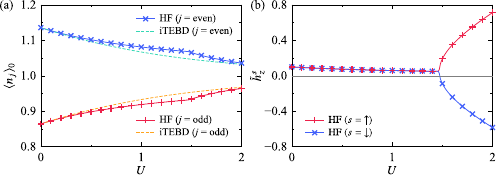}
\caption{(a)~Electron density $n_j = n_{j\uparrow} + n_{j\downarrow}$ calculated with the HF approximation (solid lines) and the iTEBD with $D = 400$ (dashed lines).
(b)~Renormalized model parameter $\tilde{h}_{z}^{s}$ in the HF approximation.
The system undergoes a second-order phase transition from a nonmagnetic ferroelectric phase to an antiferromagnetic ferroelectric phase at $U \approx 1.47$.
The parameters are set to $h_y = 0.5$ and $h_z = 0.1$ in (a) and (b).
}
\label{fig:hf}
\end{figure}

In Fig.~\ref{fig:hf}, we show the calculated $\tilde{h}_z^{s}$ in the HF approximation, along with the electron density $n_j$, as functions of the interaction strength $U$.
In the region $U \leq 0.2$ discussed in the main text, the electron density in the HF approximation [Fig.~\ref{fig:hf}(a)] agrees well with the exact iTEBD results.
In this regime, the values of $\tilde{h}_z^s$ shown in Fig.~\ref{fig:hf}(b) are only slightly smaller than the $U = 0$ value of $\tilde{h}_z^s = h_z = 0.1$, and they are identical for both spin-up and spin-down components.
Specifically, for $U = 0.2$, we find $\tilde{h}_z^{\uparrow} = \tilde{h}_z^{\downarrow} = 0.088$ and confirm that $\chi^{(2)}(\omega_1, \omega_2)$ remains qualitatively unchanged from the $U = 0$ case.
These results strongly suggest that, even though the ground states for $U \lesssim 0.2$ are well approximated by the HF theory, the observed frequency-dependent sign reversal in $\chi^{(2)}(\omega_1, \omega_2)$ (Fig.~4) is a consequence of many-body effects that are not captured by one-body physics.

